\documentclass[a4paper,11pt]{article}
\usepackage[dvips]{graphicx}
\usepackage{amssymb}
\usepackage{amsmath}
\usepackage{cite}

\begin{document}

\title{Loop Quantum Cosmology Matter Bounce Reconstruction from $F(R)$ Gravity Using an Auxiliary Field}
\author{
V.K. Oikonomou$^{1}$\,\thanks{v.k.oikonomou1979@gmail.com; voiko@physics.auth.gr}\\ \\
$^{1).}$Department of Theoretical Physics, Aristotle University of Thessaloniki,\\
54124 Thessaloniki, Greece
} \maketitle 

\begin{abstract}

Using the reconstruction technique with an auxiliary field, we investigate which $F(R)$ gravities can produce the matter bounce cosmological solutions. Owing to the specific functional form of the matter bounce Hubble parameter, the reconstruction technique leads, after some simplifications, to the same Hubble parameter as in the matter bounce scenario. Focusing the study to the large and small cosmic time $t$ limits, we were able to find which $F(R)$ gravities can generate the matter bounce Hubble parameter. In the case of small cosmic time limit, which corresponds to large curvature values, the $F(R)$ gravity is $F(R)\sim R+\alpha R^2$, which is an inflation generating gravity, and at small curvature, or equivalently, large cosmic time, the $F(R)$ gravity generating the corresponding limit of the matter bounce Hubble parameter, is $F(R)\sim \frac{1}{R}$, a gravity known to produce late-time acceleration. Thus we have the physically appealing picture in which a Jordan frame $F(R)$ gravity that imitates the matter bounce solution at large and small curvatures, can generate Starobinsky inflation and late-time acceleration. Moreover, the scale factor corresponding to the reconstruction technique coincides almost completely to the matter bounce scenario scale factor, when considered in the aforementioned limiting curvature cases. This is scrutinized in detail, in order to examine the validity of the reconstruction method in these limiting cases, and according to our analysis, exact agreement is achieved.
\end{abstract}


\section*{Introduction}

Observational data in the late 90's confirmed one of the most striking physical phenomenon in modern astrophysics and cosmology, the fact that the universe expands in an accelerating way \cite{riess}. This result initiated a research stream with aim to model in a theoretically consistent way this late-time acceleration. In addition to this late-time acceleration, confirmed by using standard candles as references, recent observational results \cite{planck,bicep} confirmed the existence of the B-mode power spectrum \cite{bicep}, thus indirectly validating the existence of an early time acceleration of the universe. One of the modern cosmology largest goals is to provide an explanation for the late-time and early time acceleration in a unified way. 

Modified theories of gravity have proved to be one of the most appealing candidate theories for the description of the acceleration eras of the universe, since they provide a theoretical framework which can be consistent with observational data coming from cosmological scales and also from astrophysical data, considering the local implications of such gravitational modifications of general relativity. The most representative candidate among the modified theories of gravity, are the $F(R)$ theories of gravity, in which the Ricci scalar appearing in the standard description of the general relativity Einstein-Hilbert action, is replaced by the more general function $F(R)$. An important stream of papers, devoted on various aspects of this vast research issue, consult \cite{reviews1,reviews2,reviews3,reviews4,reviews5,reviews8,reviews9,
importantpapers1,importantpapers2,importantpapers3,importantpapers4,importantpapers5,
importantpapers6,importantpapers8,importantpapers9,importantpapers10,importantpapers11,
importantpapers12,importantpapers13,importantpapers14,importantpapers15,importantpapers17,
importantpapers18,importantpapers19,importantpapers20}, and references therein. Among these, it worths to mention that the first consistent description of late-time and early time acceleration, within the theoretical framework of $F(R)$ theories, was given by Nojiri and Odintsov in \cite{sergeinojirimodel}. In the $F(R)$ gravities, late-time acceleration is attributed to an indirectly observable component of the energy momentum tensor, which is known as dark energy. Dark energy in $F(R)$ theories is described by a negative pressure perfect fluid which modifies the right hand side of the Einstein equations and this is the reason why it is called sometimes geometric dark energy. For alternative theories to $F(R)$ gravities that can consistently describe dark energy, the reader is referred to \cite{capo,capo1,peebles,faraonquin,tsujiintjd} and references therein. For a recent study on unified dark energy with quintessential inflation \cite{quintense}, see \cite{saridakismyrzakulov}.

The recent experimental results of Planck \cite{planck} and BICEP \cite{bicep} have restricted the viability of inflationary models, excluding models with, for example, chaotic inflation generating power law potentials, exponential potential models and inverse power law models. Moreover, these observational data seem to favor an $F(R)$ gravity with power law function, namely the $R^2$ gravity, firstly studied in the 80's and known to describe inflation \cite{starobinsky}. However, although that many inflationary potentials can be not viable in the standard cosmological contexts that these were firstly developed, they can be embedded in theories which can produce results compatible of observational data. One particularly interesting scenario for inflation is provided by the matter bounce scenario, predicted from holonomy corrected Loop Quantum Cosmology (shortened to LQC hereafter) \cite{LQC1,LQC2,LQC3,LQC4,LQC5sing,LQC6sing,LQC7sing,LQC7sing1,LQC8,LQC9,LQC10,LQC11,LQC12,LQC13,mbounce1,mbounce2,mbounce3,mbounce4,mbounce5,mbounce6,mbouncersquarefr,mbounce8,mbounce9,mbounce10,mbounce11}.
For an informative but incomplete list on this subject, the reader is referred to \cite{LQC1,LQC2,LQC3,LQC4,LQC5sing,LQC6sing,LQC7sing,LQC7sing1,LQC8,LQC9,LQC10,LQC11,LQC12,LQC13,mbounce1,mbounce2,mbounce3,mbounce4,mbounce5,mbounce6,mbouncersquarefr,mbounce8,mbounce9,mbounce10,mbounce11} and references therein. See also reference \cite{myrza} for bounce solutions in viscous cosmology scenario. Two of the appealing features of the LQC matter bounce scenario are that, firstly, singularities are elegantly and consistently resolved \cite{LQC5sing,LQC6sing,LQC7sing,LQC7sing1} and secondly, concordance with the observational data of Planck is feasible \cite{mbounce1,mbounce2,mbounce3,mbounce4,mbounce5,mbounce6,mbouncersquarefr,mbounce8,mbounce9,mbounce10,mbounce11}.

Owing to the appealing properties that the matter bounce scenario of holonomy corrected LQC brings along, we shall investigate from which $F(R)$ gravity the cosmological solutions of the matter bounce scenario can be generated in the context of modified gravity theories. Particularly, using the metric formalism of $F(R)$ theories of gravity \cite{reviews1,reviews2,reviews3,reviews4,reviews5,reviews8,reviews9}, and assuming a flat Friedmann-Robertson-Walker (shortened to FRW hereafter) metric, we shall use a very well known reconstruction technique \cite{importantpapers3} in order to find which $F(R)$ gravity can reproduce the cosmology described by the matter bounce scenario. In the technique we shall use, an auxiliary scalar field is used in order to obtain the reconstructed gravities, in contrast to another reconstruction method \cite{importantpapers12}, which is more direct than the present technique. For a recent approach adopting the latter technique, in order to reconstruct $F(R)$ gravities that reproduce the matter bounce cosmological solutions, see \cite{sergeioikonomou}. In principle, deviations between the two methods are expected due to the different approach to the problem, an issue we intent to explain in a formal way in a later section in this paper. The matter content of the $F(R)$ gravities shall also be taken into account, with matter consisting of collision-less, non relativistic matter and also relativistic matter. The cosmology to be reconstructed is the matter bounce cosmology, expressed in terms of the matter bounce Hubble parameter, which will be the starting point of our analysis. In order to obtain analytic solutions to the problem, we shall focus our study in two extreme limits of the curvature, that is, in the small and large curvature limits. As we shall see the results are particularly interesting since in the small curvature limit, the $F(R)$ gravity that reproduces the large $R$ limit of the matter bounce cosmology, is given by $F(R)\sim \frac{1}{R}$. Recall that the small curvature limit corresponds to the late cosmic time era, and the aforementioned function is known to produce late time acceleration, which is a rather appealing feature. On the other hand, in the large curvature limit, which corresponds to early cosmic time, the $F(R)$ gravity that reproduces the large curvature limit of the matter bounce cosmology is, at leading order, given by the function $F(R)\sim R+\alpha R^2$. The latter gravity is known to produce Starobinsky inflation \cite{reviews1,reviews2,reviews3,reviews4,reviews5,reviews8,reviews9,starobinsky}. Having an $R^2$ gravity at hand, with a parameter $\alpha$ that is free to vary, we briefly study the very well known inflation properties of the aforementioned gravity \cite{sergeistarobinsky}, and also we recall, following the relevant literature on this \cite{mbouncersquarefr}, how this modified gravity can be embedded in a LQC framework by introducing holonomy corrections in the $F(R)$ gravity explicitly. After briefly recalling these two important theoretical attributes of the $R^2$ gravity, we present a mathematical reasoning with regards to the reconstruction method that we use. We have to make an important notice with regards to the auxiliary field augmented method we shall use here, that in the case at hand, that is matter bounce cosmology, the method is applicable as we can see, owing to the particular form of the Hubble parameter. This fact as we shall see justifies our approach as we will explicitly demonstrate, and renders the result credible, from a mathematical point of view at least.

We need to note that there exist in the literature quite important works, in which quite similar to the present work issues. For example, in \cite{ref1}, where the reconstruction of the LQC dynamics in the context of Palatini $F(R)$ theory coupled to a massless scalar field was carried out for the first time. In the same work, the authors provided a framework towards the understanding the continuum properties of the underlying discrete structure of the quantum geometry and also provided some insights for the extraction of physical predictions, such as those related to gauge invariant cosmological perturbations. In addition, in \cite{ref2} the authors provided an inverse reconstruction method using Supernova Type Ia data, so that they were able to reconstruct Palatini $F(R)$ gravity theories and in \cite{ref3}, by using the technique of order reduction, the author was able to derive the second order field equations that provided only the physical degrees of freedom of an LQC $F(R)$ theory in the metric formalism. Moreover, in \cite{ref4}, working in the Palatini formalism, the authors have shown that it is possible to reconstruct the LQC dynamics for arbitrary density dependence, using an $F(R)$ scenario, with a non-trivial matter-curvature coupling. This work \cite{ref4}, relaxes an unnatural constraint imposed in \cite{ref5}, in order to solve the same problem in the metric formalism. Noteworthy is the fact the works \cite{ref4,ref5} are quite relevant, since in principle on is not obliged to specify the time-dependence of the energy density function appearing on the right-hand side of the LQC effective equations. Notice that in the present work we shall use the metric formalism approach and we shall investigate which vacuum $F(R)$ gravity can produce the matter bounce scale factor, with vacuum meaning that no matter fluids are present.

This paper is organized as follows: In section 1 we present some fundamental properties of $F(R)$ theories of gravity in the Jordan frame. In addition we briefly introduce the matter bounce cosmological solutions of the holonomy corrected LQC and explicitly apply and work out the reconstruction technique using an auxiliary scalar field. We study the problem in the small and large cosmic time limits, in the presence of matter fields, both relativistic and non-relativistic. Moreover we very briefly present the inflation study of the $R^2$ gravity in the Einstein frame and following the relevant literature, we describe how the embedding of the $R^2$ gravity in a LQC context can be done explicitly. In section 2, we investigate the validity of our approximation presenting exact values of the measured quantities to further support our approximations. Furthermore, we examine whether the scale factor of the matter bounce scenario can be produced within our approximate techniques, and up to which extent this occurs. Also we study whether there can be any overlap of our reconstructed gravities with other bounce generating reconstructed $F(R)$ gravities studied extensively in the relevant literature. In section 3, we present in a mathematically formal way the details of the reconstruction method we used. The conclusions follow in the end of the paper. Finally, in the appendix A we present details of the solutions we found in the text.

\section{LQC Cosmological Bounce Solutions and Auxiliary Field Reconstruction Method}

Before starting off with the main subject of this article, it worths recalling some basic features of $F(R)$ modified theories of gravity, in the Jordan frame. The reader is referred to the review articles \cite{reviews1,reviews2,reviews3,reviews4,reviews5,importantpapers1,
importantpapers2,importantpapers3,importantpapers4,importantpapers5,importantpapers6,importantpapers8,importantpapers9,importantpapers10,importantpapers11,importantpapers12,importantpapers13}
for a thorough and detailed analysis on these issues.
 
The Jordan frame action of $F(R)$ modified theories of gravity in four dimensions is:
\begin{equation}\label{action}
\mathcal{S}=\frac{1}{2\kappa^2}\int \mathrm{d}^4x\sqrt{-g}F(R)+S_m(g_{\mu \nu},\Psi_m),
\end{equation}
with the parameter $\kappa$ being related to the gravitational constant according to the relation, $\kappa^2=8\pi G$ and additionally, $S_m$ is the matter action describing all the matter fields $\Psi_m$.

In the metric formalism, the equations of motion are obtained by varying the action (\ref{action}) with respect to the metric $g_{\mu \nu}$, and in this way, the equations of motion are the following,
\begin{align}\label{modifiedeinsteineqns}
R_{\mu \nu}-\frac{1}{2}Rg_{\mu \nu}=\frac{\kappa^2}{F'(R)}\Big{(}T_{\mu
\nu}+\frac{1}{\kappa^2}\Big{(}\frac{F(R)-RF'(R)}{2}g_{\mu \nu}+\nabla_{\mu}\nabla_{\nu}F'(R)-g_{\mu
\nu}\square F'(R)\Big{)}\Big{)}.
\end{align}
In this particular case, the prime on the $F(R)$ function expresses differentiation with respect to the Ricci scalar $R$, that is $F'(R)=\partial F(R)/\partial R$, and also $T_{\mu \nu}$ denotes the energy momentum tensor corresponding to ordinary matter fields. The energy momentum tensor receives an additional contribution coming from the $F(R)$ sector only. Indeed, it is this latter feature that makes the $F(R)$ theories of gravity, modified with regards to ordinary general relativity. As we can see, the left hand side is the same as in general relativity, but in the right hand side, if it is assumed that it originates from some sort of exotic matter, the new energy momentum tensor can be written in the following way:
\begin{equation}\label{newenrgymom}
T^{eff}_{\mu \nu}=\frac{1}{\kappa}\Big{(}\frac{F(R)-RF'(R)}{2}g_{\mu
\nu}+\nabla_{\mu}\nabla_{\nu}F'(R)-g_{\mu \nu}\square F'(R)\Big{)}.
\end{equation}
This new energy momentum tensor (\ref{newenrgymom}) is absent from the Einstein-Hilbert gravity, and some parts of it can explicitly model the dark energy in $F(R)$ theories of modified gravity. This is actually how the terminology geometric dark energy originates from. Before closing this overview, it worths presenting the metric we shall use in the following sections, which is a flat FRW spacetime being described by the following metric,
\begin{equation}\label{metricformfrwhjkh}
\mathrm{d}s^2=-\mathrm{d}t^2+a^2(t)\sum_i\mathrm{d}x_i^2
\end{equation}
The Ricci scalar in this metric background equals to:
\begin{equation}\label{ricciscal}
R=6(2H^2+\dot{H}),
\end{equation}
with $H(t)$ the Hubble parameter and with the ``dot'' it is indicated differentiation with respect to the cosmological time $t$.

\subsection{Applying the Reconstruction Method}

In the context of LQC, consistency is ensured by making specific modifications to the operators describing the Hamiltonian system \cite{mbounce1,mbounce2,mbounce3,mbounce4,mbounce5,mbounce6}. In addition to this, and owing to holonomy corrections, the Friedmann equation is modified and particularly in the context of LQC for a matter dominated universe, it is given by \cite{mbounce10,mbounce11}:
\begin{equation}\label{holcor1}
H^2=\frac{\rho}{3}\left (1-\frac{\rho}{\rho_c}\right ),{\,}{\,}{\,}\dot{\rho}(t)=-3H\rho(t)
\end{equation}
with the matter-energy density being equal to, 
\begin{equation}\label{rhosol}
\rho (t)=\frac{\rho_c}{\frac{3}{4}t^2+1}
\end{equation}
Solving (\ref{holcor1}) and having in mind (\ref{rhosol}) we get the following solutions, with regards to the Hubble parameter $H(t)$ and for the scale factor $a(t)$ in the matter bounce scenario of LQC \cite{mbounce10,mbounce11}:
\begin{equation}\label{holcorrLQCsol}
a(t)=\left (\frac{3}{4}\rho_ct^2+1\right )^{1/3},{\,}{\,}{\,}H(t)=\frac{\frac{1}{2}\rho_ct}{\frac{3}{4}\rho_ct^2+1}
\end{equation}
The Hubble parameter appearing in the equation above  (\ref{holcorrLQCsol}), will be our starting point of the reconstruction technique we shall deploy. This is a crucial point in our analysis, and it worths describing this point in detail. There are two reconstruction techniques in the literature \cite{importantpapers3,importantpapers12}, with the one developed in \cite{importantpapers12} being more restrictive in the vector space of solutions. With regards to this, in the discussion section 3, we discuss this from a mathematical perspective. The method we shall use here, was firstly developed in \cite{importantpapers3} and results to a larger family of solutions, because the scale factor is assumed to be of a specific form and also certain simplifications are done in order the problem is solved. Practically these simplifications are some sort of perturbation theory at the level of Euler-Lagrange equations. This kind of delicate mathematical issue, shall be thoroughly and rigidly addressed from a mathematical point of view in section 3 of the present paper, since there is much more to say on this. But we have to note that the exact form of the scale factor appearing in relation (\ref{holcorrLQCsol}) cannot be realized exactly within the theoretical framework we are going to use in this paper, only the Hubble parameter, and this is due to the fact that the scale factor is assumed to be of a specific from, which severely broadens the family of solutions. Nevertheless in the large and small cosmic time limits, we have almost absolute concordance between the scale factor obtained by our method and the matter bounce scale factor. The action of  the general $F(R)$ gravity is given by,
\begin{equation}\label{action1dse}
\mathcal{S}=\frac{1}{2\kappa^2}\int \mathrm{d}^4x\sqrt{-g}F(R)+S_m(g_{\mu \nu},\Psi_m),
\end{equation}
and the first FRW equation appearing in relation (\ref{modifiedeinsteineqns}) can be written as:
\begin{equation}\label{frwf1}
-18\left ( 4H(t)^2\dot{H}(t)+H(t)\ddot{H}(t)\right )F''(R)+3\left (H^2(t)+\dot{H}(t) \right )F'(R)-\frac{F(R)}{2}+\kappa^2\rho=0
\end{equation}
with $F'(R)=\frac{\mathrm{d}F(R)}{\mathrm{d}R}$ and the Ricci scalar $R$ given by relation (\ref{ricciscal}) as a function of the time variable. The reconstruction technique developed in \cite{importantpapers3} can be used to define an implicit form of an $F(R)$ function, given the general form of the Hubble parameter as a function of the cosmic time $t$. It worths recalling the technique, since this is essential for our analysis. For further details, the reader is referred to consult reference \cite{importantpapers3}, where this technique was developed for the first time. Note that it is possible to find the purely geometrical $F(R)$ gravity without taking into account matter fluids, but it is also possible to include matter fluids in the process. We shall include matter fluids with general cosmological parameters $w_i$, assuming that there are ''$i$'' in number species of matter. Using an auxiliary scalar field $\phi $, the action of relation (\ref{action}) for the $F(R)$ gravity can be rewritten in the following form,
\begin{equation}\label{neweqn123}
S=\int \mathrm{d}^4x\sqrt{-g}\left ( P(\phi )R+Q(\phi ) +\mathcal{L}_{mat} \right )
\end{equation}
with the term $\mathcal{L}_m$ describing the matter Lagrangian entering the matter action in the usual way,
\begin{equation}\label{matterlagra}
S_{m}=\int \mathrm{d}^4x\sqrt{-g}\mathcal{L}_{mat}
\end{equation}
The functions $P(\phi )$ and $Q(\phi )$ are what actually will provide us with the final form of the $F(R)$ gravity as we shall shortly see. The scalar field $\phi $ is an auxiliary dynamical (depends on time) degree of freedom, which can be easily seen due to the lack of a kinetic term. Variation of the action (\ref{neweqn123}) with respect to this auxiliary degree of freedom, yields the following equation,
\begin{equation}\label{auxiliaryeqns}
P'(\phi )R+Q'(\phi )=0
\end{equation}
which can be solved with respect to $\phi $, to yield the functional dependence of $\phi (R)$, as a function of $R$. In relation (\ref{auxiliaryeqns}), the prime denotes as usually does, differentiation with respect to $\phi$. Then the $F(R)$ gravity is obtained from $\phi (R)$ by substituting it to the original $F(R)$ action, that is,
\begin{equation}\label{r1}
F(\phi( R))= P (\phi (R))R+Q (\phi (R))
\end{equation}
So practically having an equation for $P(\phi )$ and $Q(\phi )$ will yield us the $F(R)$ gravity. This differential equation in terms of $P(\phi )$ and $Q(\phi )$ can be derived by varying equation (\ref{neweqn123}) with respect to the metric tensor, and applying these for a spatially flat FRW universe with scale factor $a(t)$ and Hubble parameter $H(t)$. Doing that we end up with the following differential equation,
\begin{align}\label{r2}
& -6H^2P(\phi (t))-Q(\phi (t) )-6H\frac{\mathrm{d}P\left (\phi (t)\right )}{\mathrm{d}t}+\rho_i=0 \\ \notag &
\left ( 4\dot{H}+6H^2 \right ) P(\phi (t))+Q(\phi (t) )+2\frac{\mathrm{d}^2P(\phi (t))}{\mathrm {d}t^2}+\frac{\mathrm{d}P(\phi (t))}{\mathrm{d}t}+p_i=0
\end{align}
which, by eliminating $Q(\phi (t))$ one gets,
\begin{equation}\label{r3}
2\frac{\mathrm{d}^2P(\phi (t))}{\mathrm {d}t^2}-2H(t) P(\phi (t))+4\dot{H}\frac{\mathrm{d}P(\phi (t))}{\mathrm{d}t}+\rho_i+p_i=0
\end{equation}
with $\rho_i,p_i$ denoting the mass-energy density and pressure of the matter fluid present, respectively. It is proved in reference \cite{importantpapers3} that due to the equivalence of actions (\ref{action}) and (\ref{neweqn123}), the scalar field may be redefined to be equal to the cosmic time $t$, that is $\phi =t$ (see appendix of \cite{importantpapers3}, for a proof in the case of power law functions). Assuming conventional matter fluids with constant cosmological equation of state parameter $w_i$ and also most importantly that the scale factor is written in the form,
\begin{equation}\label{r4}
a=a_0e^{g(t)}
\end{equation}
with $a_0$ constant, the differential equation (\ref{r3}) for $P(\phi )$ can be written in the following form,
\begin{align}\label{r5}
& 2\frac{\mathrm{d}^2P(\phi (t))}{\mathrm {d}t^2}-2g'(\phi )\frac{\mathrm{d}P(\phi (t))}{\mathrm{d}t}+4g''(\phi ) P(\phi (t))+\sum_i(1+w_i)\rho_{i0}a_0^{-3(1+w_i)}e^{-3(1+w_i)g(\phi )}=0
\end{align}
This form of the scale factor, given in relation (\ref{r4}), renders the technique an approximate method, which in the extreme limits of cosmic time, is an exact method giving results that coincide with the matter bounce solutions. It is a simple task to directly find $Q(\phi )$, given $P(\phi )$. Indeed, using relation (\ref{r2}), $Q(\phi )$ is equal to,
\begin{equation}\label{r5a}
Q(\phi )=-6g'(\phi )^2P(\phi )-6g'(\phi )\frac{\mathrm{d}P(\phi )}{\mathrm{d}\phi }+\sum_i(1+w_i)\rho_{i0}a_0^{-3(1+w_i)}e^{-3(1+w_i)g(\phi )}
\end{equation}
The key point of the method is to find the relation that connects the Ricci scalar $R$ with $\phi $, given the cosmological evolution of the universe in terms of a specific Hubble parameter or equivalently the scale factor. Then, by substituting the relation $\phi =\phi (R)$, one can easily substitute this in the differential equation (\ref{r5}) in order to find the specific $P(\phi (R))$ and $Q(\phi (R))$ and thereby, the $F(R)$ gravity that generates the specific cosmological evolution. 

We shall promptly apply this technique to find which pure or matter containing $F(R)$ gravity generates the matter bounce solutions of LQC, given in relation (\ref{holcorrLQCsol}). This is practically our direct link to the matter bounce cosmology, the Hubble parameter. Let us see how this method works. Since we have assumed the particular form of relation (\ref{r4}), it is rather difficult to reproduce the scale factor of the LQC bounce solutions, without taking some limiting case. As we shall see, our results can be given in a closed form owing to the particular form of the Hubble parameter $H(t)$ corresponding to the matter bounce solution. Particularly, the matter bounce Hubble parameter (\ref{holcorrLQCsol}) can be written as follows,
\begin{equation}\label{r6}
H(t)=\frac{\frac{1}{2}\rho_ct}{\frac{3}{4}\rho_ct^2+1}=\frac{h(t)}{t}
\end{equation}
with $h(t)$ being equal to,
\begin{equation}\label{r7}
h(t)=\frac{h_fqt^2}{1+qt^2}
\end{equation}
and in addition $h_f$, $q$ stand for,
\begin{equation}\label{r7a}
h_f=\frac{2}{3},{\,}{\,}{\,}q=\frac{3\rho_c}{4}
\end{equation}
This particular form of the Hubble parameter, namely relation (\ref{r6}), is crucial to our analysis. Note that $h(t)$ is a slowly varying function of time, and this is what makes this particular case so interesting. Recall that a slowly varying function of $t$ is defined to be a function that can take the following form,
\begin{equation}\label{r8}
h(t)=c(t)e^{\int_{t_0}^t\frac{y(x)}{x}}\mathrm{d}x
\end{equation}
with $c(t)$ a measurable non-negative function of $t$, and $c(t)$ and $y(x)$ are defined in such a way, so that the following requirements are met:
\begin{equation}\label{r9}
\lim_{t\rightarrow \infty}c(t)=c_0,{\,}{\,}{\,}\lim_{x\rightarrow \infty }y(x)\rightarrow 0
\end{equation}
with $c_0$ a finite number. As a consequence, the function $h(t)$ satisfies $\forall $ $z$ $\in $ $\mathbb{R}$, the following relation,
\begin{equation}\label{r10}
\lim_{t\rightarrow \infty}\frac{h(zt)}{h(t)}=1
\end{equation}
It can be easily verified that the function (\ref{r8}) satisfies the requirement (\ref{r10}), so it is slowly varying. As a consequence of this, we assume that the function $g(t)$ appearing in the scale factor in relation (\ref{r4}) can take the following form,
\begin{equation}\label{r11}
g(\phi )=h(\phi)\ln \left( \frac{\phi}{\phi_0}\right )
\end{equation}
with $\phi_0$ some constant to be determined shortly. Notice that we used $\phi $ instead of $t$, but $\phi=t$ so we continuously use this interplay of these variables. In addition, owing to the fact that $h(\phi )$ is slowly varying, it's derivatives can be neglected, which can be easily verified from the exponential representation of the slowly varying function, namely relation (\ref{r9}). Having the exact form of the function $g(\phi )$ and of the Hubble parameter $H(t)$, we can promptly solve the differential equation (\ref{r5}). The Hubble parameter corresponding to the scale factor (\ref{r4}), with the function $g(t)$ given by (\ref{r11}), is equal to,
\begin{equation}\label{r11a}
H(t)=\frac{h(t)}{t}+h'(t)\ln\left (\frac{t}{t_0} \right ) 
\end{equation} 
Since $h(t)$ is slowly varying, the derivative $h'(t)$ can be ignored, and hence the Hubble parameter of our approximation method reads,
\begin{equation}\label{r11b}
H(t)\simeq \frac{h(t)}{t}
\end{equation}
which is identical to the matter bounce Hubble parameter. This coincidence is crucial for our reconstruction technique, since we shall find the $F(R)$ gravities that can reproduce this Hubble parameter. The capability of our approximate method to reproduce the Hubble parameter and also the scale factor of the matter bounce scenario, shall be critically investigated in detail in a later section. 

Owing to the slow varying behavior of $h(t)$, by ignoring the higher derivatives $h'(t),h''(t)$, and by using the function $g(t)$ as given in (\ref{r11}), the differential equation (\ref{r5}) becomes,
\begin{align}\label{r12}
& 2\frac{\mathrm{d}^2P(\phi (t))}{\mathrm {d}t^2}-\frac{h(\phi )}{\phi }\frac{\mathrm{d}P(\phi (t))}{\mathrm{d}t}-\frac{2h(\phi )}{\phi^2} P(\phi (t))
\\ \notag & +\sum_i(1+w_i)\rho_{i0}a_0^{-3(1+w_i)}\left ( \frac{\phi }{\phi_0}\right )^{-3(1+w_i)h(\phi )}=0
\end{align} 
We shall present the solution of the above differential equation later, since we first need to find the functional dependence of $\phi$ as a function of $R$, corresponding to the Hubble parameter (\ref{r7}). In order to do so, we calculate the Ricci scalar $R(\phi )$, using relations (\ref{ricciscal}) and (\ref{r6}) and (\ref{r7}), and neglecting the higher order derivatives for the slowly varying function $h(t)$, we get,
\begin{equation}\label{r13}
R(\phi )\simeq \frac{6\left (-h(\phi )+2h(\phi )^2  \right )}{\phi^2}
\end{equation}
We aim to solve the above equation with respect to $\phi^2$. Setting $u=\phi^2$, we get the following third order polynomial of the variable $u$,
\begin{equation}\label{r14}
q^2Ru^3+\left ( 2qR-6q^2+h_fq^2-2h_f^2q^2\right )u^2+\left ( R-12q+h_fq \right )u-6=0 
\end{equation}
The real solution to this equation with respect to $u=\phi^2$ is of the following form,
\begin{align}\label{r15}
& u = \phi^2 =-\frac{2}{3 q}+\frac{2}{R}-\frac{h_f}{3 R}+\frac{2 h_f^2}{3 R}\\ \notag &
+ \Big{(}4\times 2^{1/3} q+\frac{12\times 2^{1/3} q^2}{R}+\frac{2^{1/3} R}{3}+\frac{2^{1/3}q h_f}{3}\\ \notag & -\frac{4\times 2^{1/3} q^2 h_f}{R}- \frac{8\times 2^{1/3} q h_f^2}{3}+\frac{25\times 2^{1/3} q^2 h_f^2}{3R}-\frac{4\times 2^{1/3} q^2 h_f^3}{3R}+4\times 2^{1/3} q^2 h_f^4+\frac{1}{3 2^{1/3} q^2 R}\Big{)} \\ \notag &
\times \frac{1}{\left ( \alpha_0+\alpha_1R+\alpha_2R^2+\alpha_3R^3 +\sqrt{\beta_2R^2+\beta_3 R^3+\beta_4 R^4+\beta_5R^5} \right )^{1/3}}
\end{align}
where the coefficients $\alpha_i,\beta_i$ are given in the appendix A for the readers convenience. Having the solution (\ref{r15}) at hand, which give the explicit form of the auxiliary field $\phi $ as a function of the Ricci scalar, we can proceed by solving the differential equation (\ref{r12}) and express the functions $P(\phi(R))$ and $Q(\phi(R))$ as functions of the Ricci scalar and by substituting the results to equation (\ref{r1}) we will find the reconstructed $F(R)$ gravity that produces the matter bounce cosmological evolution. In order to simplify the investigation, we shall study the solutions in the large $R$ and small $R$, which correspond to small $t$ and large $t$ limits of the theory respectively. Before going to this, it worths to give here the general solution of the differential equation (\ref{r12}), and also the expression for $Q(\phi )$ in order to directly apply the results for the large and small $R$ limits. So the general solution for the differential equation (\ref{r12}) is of the following form \cite{importantpapers3},
\begin{align}\label{r16}
& P(\phi )=c_1\phi^{\frac{h(\phi )-1+\sqrt{h(\phi)^2+6h(\phi )+1}}{2}}+c_2\phi^{\frac{h(\phi )-1-\sqrt{h(\phi)^2+6h(\phi )+1}}{2}}+\sum_iS_i(\phi )\phi^{-3(1+w_i)h(\phi )+2}
\end{align}
with the coefficient $S_i(\phi )$ being equal to,
\begin{align}\label{r17}
& S_i(\phi )=-\left ((1+w_i)\rho_{i0}a_0^{-3(1+w_i)}\phi_0^{3(1+w)h(\phi )} \right )\\ \notag & \left ( 6(1+w_i)(4+3w_i)h(\phi)^2-2(13+9w)h(\phi)+4\right )^{-1}
\end{align}
Having the general form of the solution (\ref{r16}), we may find the function $Q(\phi)$ by substituting (\ref{r11}) and (\ref{r17}) in equation (\ref{r5a}), and $Q(\phi )$ reads,
\begin{align}\label{r18}
& Q(\phi )=-6h(\phi )c_1\left (h(\phi )+\frac{h(\phi )-1+\sqrt{h(\phi)^2+6h(\phi )+1}}{2}\right )\phi^{\frac{h(\phi )-1+\sqrt{h(\phi)^2+6h(\phi )+1}}{2}-2}\\ \notag &
-6h(\phi )c_2\left (h(\phi )+\frac{h(\phi )-1-\sqrt{h(\phi)^2+6h(\phi )+1}}{2}\right )\phi^{\frac{h(\phi )-1+\sqrt{h(\phi)^2+6h(\phi )+1}}{2}-2}\\ \notag & +\sum_i\left ( -6h(\phi )(-2(2+3w)h(\phi )+2+S_i(\phi )+\rho_{i0}a_0^{-3(1+w_i)\phi_0^{3(1+w_i)h(\phi)}})\right )\phi^{-3(1+w_i)h(\phi )}
\end{align}

\subsection{$F(R)$ Gravity in the Large $R$ Limit}

Let us first study the large $R$ limit of the theory, which actually describes early times when the curvature of the universe has large values. The solution (\ref{r15}) for $\phi^2$, in the large $R$ limit, reads,
\begin{align}\label{r16a}
& \phi^2\simeq -\frac{2}{3 q}+ \Big{(}\frac{2^{1/3} R}{3}+\mathcal{A}_1\Big{)}\frac{1}{\alpha_5R}
\end{align}
which is equal to,
\begin{align}\label{r17b}
& \phi^2\simeq \frac{\mathcal{A}_1}{\alpha_5R}
\end{align}
so finally,
\begin{equation}\label{r17a}
\phi \sim (\mathcal{A}_1)^{1/2}\alpha_5^{-1/2}R^{-1/2}
\end{equation}
where we have set $\mathcal{A}_1$ to be equal to,
\begin{align}\label{r18a}
\mathcal{A}_1=4\times 2^{1/3} q+\frac{2^{1/3}q h_f}{3}- \frac{8\times 2^{1/3} q h_f^2}{3}+4\times 2^{1/3} q^2 h_f^4
\end{align}
and $\alpha_5$ can be found in appendix A. We can promptly find the function $P(\phi )$ as a function of $R$ in the large $R$ limit by substituting (\ref{r17a}) in (\ref{r16}). It would be convenient however to find simplify the expression (\ref{r16}) in the large $R$ limit, or equivalently in the small $t$ (or small $\phi $ limit, recall $\phi=t$). The function $h(t)$ has the following two limiting values, with respect to $t$,
\begin{equation}\label{r19}
\lim_{t\rightarrow 0}h(t)=0,{\,}{\,}{\,}\lim_{t\rightarrow \infty }h(t)=h_f
\end{equation} 
What actually interests us in this section is the small $t$ limit, owing to the fact that $\lim_{t\rightarrow \infty}h(t)=0$. This fact simplifies the final solution for $P(\phi )$ and $Q(\phi )$, so for the moment we shall assume the presence of various matter fields. As we shall see, they play no role in the final form of the solution. In the small $t$ limit (large $R$), the function $P(\phi) $ reads,
\begin{equation}\label{r20}
P(\phi )=c_1+c_2\phi^{-2}+\sum_iS_i\phi^2
\end{equation}
which after using (\ref{r17}), equation (\ref{r20}) becomes,
\begin{equation}\label{r21}
P(\phi (R))=c_1+\frac{c_2a_5}{\mathcal{A}_1}R+\sum_iS_i(0)\frac{\mathcal{A}_1}{\alpha_5R}
\end{equation}
so by keeping only the dominating terms in the large $R$ limit, we get,
\begin{equation}\label{r22}
P(\phi (R))\simeq c_1+\frac{c_2a_5}{\mathcal{A}_1}R
\end{equation}
Accordingly, the function $Q(\phi (R))$ reads,
\begin{equation}\label{r23}
Q(\phi(R))=\sum_iS_i(\phi )a_0^{-3(1+w_i)}
\end{equation}
Finally, in view of relations (\ref{r22}) and (\ref{r23}), the reconstructed $F(R)$ gravity of relation (\ref{r1}) in the presence of matter fluids reads,
\begin{equation}\label{r24}
F(R)\simeq \left( c_1+\sum_iS_i(\phi )a_0^{-3(1+w_i)}\right ) R+\frac{c_2a_5}{\mathcal{A}_1}R^2
\end{equation}
By choosing the coefficient of $R$ in the above expression to be equal to one,
\begin{equation}\label{r24a}
c_1+\sum_iS_i(\phi )a_0^{-3(1+w_i)}=1,{\,}{\,}{\,}
\end{equation}
and defining $\alpha$ as follows,
\begin{equation}\label{r24b}
\alpha =\frac{c_2a_5}{\mathcal{A}_1}
\end{equation}
the final form of the reconstructed $F(R)$ gravity in the large $R$ limit, is the following,
\begin{equation}\label{r24c}
F(R)\simeq R+\alpha R^2
\end{equation} 
This $F(R)$ gravity is of particular importance, since it is one candidate for inflation and currently has applications in many theoretical cosmology frameworks \cite{mbounce4,sergeistarobinsky}. Let us summarize how we ended up to this reconstructed gravity. Having started from the exact form of the Hubble parameter corresponding to matter bounce LQC solutions, using the method of reconstruction developed in \cite{importantpapers3}, using an approximation that relies on the particular form of the Hubble parameter (\ref{r6}) and therefore disregarding the derivatives of the function $h(t)$, in the high curvature (small $t$) limit, the reconstructed gravity that reproduces the Hubble parameter (\ref{r6}) is the one that gives the Starobinsky inflation \cite{reviews1,reviews2,sergeistarobinsky}. So within the slow varying function approximation, the reconstruction method yields a particularly interesting $F(R)$ gravity describing exactly the inflationary era, that is, at early cosmological times $t$.

\subsection{$F(R)$ Gravity in the Small $R$ Limit}

In the small $R$ limit, which is actually equivalent to the large $t$ (or $\phi$ since $t=\phi$ in our approximation) limit, the slow varying function $h(\phi)$ is approximately equal to $h_f$. Recall that the form of $H(t)$ dictates that $h_f=2/3$ and also $q=3\rho_c/4$. In the high $R$ limit, we have that,
\begin{equation}\label{r25}
\phi^2\simeq \frac{\mathcal{A}_2}{R}
\end{equation}
with the parameter $\mathcal{A}_2$ being equal to,
\begin{equation}\label{r26}
\mathcal{A}_2=\left ( 25\times 2^{1/3} q^2 h_f^3+ 12\times 2^{1/3} q^2 + \frac{1}{3\times 2^{1/3} q^2 }- 4\times 2^{1/3} q^2 h_f-\frac{4\times 2^{1/3} q^2 h_f^2}{3}   \right )\frac{1}{a_0^{1/3}}
\end{equation}
Therefore the function $P(\phi )$ in terms of $\phi $ in the high $\phi$ limit reads,
\begin{equation}\label{r27}
P(\phi )\simeq c_1\phi^{h_f-1+\sqrt{h_f^2+6h_f+1}}+c_2\phi^{h_f-1-\sqrt{h_f^2+6h_f+1}}+\sum_iS_i(\infty )\phi^{-3 (1+w_i)h_f+2}
\end{equation}
 or equivalently in terms of the Ricci scalar, by using relation (\ref{r25}) it can be written as follows,
 \begin{align}\label{r28}
 P(\phi (R))\simeq c_1 &\mathcal{A}_2^{\frac{h_f-1+\sqrt{h_f^2+6h_f+1}}{2}}R^{\frac{-h_f+1-\sqrt{h_f^2+6h_f+1}}{2}}+c_2\mathcal{A}_2^{\frac{h_f-1-\sqrt{h_f^2+6h_f+1}}{2}}R^{\frac{-h_f+1+\sqrt{h_f^2+6h_f+1}}{2}}
\\ \notag &+\sum_iS_i(\infty )\mathcal{A}_2^{\frac{-3(1+w_i)h_f+2}{2}}R^{\frac{3(1+w_i)h_f-2}{2}}  
\end{align}
So by keeping only the dominant terms from the above expression (\ref{r28}), in the $R\rightarrow 0$ limit, the function $P(\phi (R))$ reads,
\begin{equation}\label{r29}
P(\phi )\simeq c_1 R^{-\delta}
\end{equation}
with the parameter $\delta $ being equal to,
\begin{equation}\label{r30}
\delta =\frac{h_f-1+\sqrt{h_f^2+6h_f+1}}{2}=1
\end{equation}
Note that we made use of the fact that $h_f=2/3$ and also that the exact values of the exponents are equal to,
\begin{equation}\label{r31}
\frac{-h_f+1-\sqrt{h_f^2+6h_f+1}}{2}= -1,{\,}{\,}{\,}\frac{-h_f+1+\sqrt{h_f^2+6h_f+1}}{2}\simeq 1.33
\end{equation}
In addition, for non-relativistic matter, the exponent of $R$ is equal to,
\begin{equation}\label{r32}
\frac{3(1+w_i)h_f-2}{2}= 0
\end{equation}
while for relativistic matter, the corresponding exponent is approximately equal to,
\begin{equation}\label{r33}
\frac{3(1+w_i)h_f-2}{2}= -0.33
\end{equation}
This is why we kept only the term appearing in (\ref{r29}). In the same vain, the function $Q(\phi )$ in the large $\phi$ limit is equal to,
\begin{align}\label{r34}
&Q(\phi )\simeq -6h_fc_3(2h_f-1+\sqrt{h_f^2+6h_f+1})(\phi^2)^{\frac{h_f-1+\sqrt{h_f^2+6h_f+1}}{2}}
\\ \notag &-6h_fc_4(2h_f-1-\sqrt{h_f^2+6h_f+1})(\phi^2)^{\frac{h_f-1-\sqrt{h_f^2+6h_f+1}}{2}}
\\ \notag & +\sum_i S_i(\infty )(-6h_f (2+3w_i)h_f+2)+\rho_{i0}a_0^{-3{1+w_i}}\phi_0^{3(1+w_i)h_f}(\phi^2)^{-\frac{3(1+w_i)h_f}{2}}
\end{align}
so in terms of the Ricci scalar $R$ we get the following expression for $Q(\phi (R) )$,
\begin{align}\label{r35}
&Q(\phi (R) )\simeq -6h_fc_3(2h_f-1+\sqrt{h_f^2+6h_f+1})\mathcal{A}_2^{\frac{h_f-1+\sqrt{h_f^2+6h_f+1}}{2}}R^{\frac{-h_f+1-\sqrt{h_f^2+6h_f+1}}{2}}
\\ \notag &-6h_fc_4(2h_f-1-\sqrt{h_f^2+6h_f+1})\mathcal{A}_2^{\frac{h_f-1-\sqrt{h_f^2+6h_f+1}}{2}}R^{\frac{-h_f+1+\sqrt{h_f^2+6h_f+1}}{2}}
\\ \notag & +\sum_i S_i(\infty )(-6h_f (2+3w_i)h_f+2)+\rho_{i0}a_0^{-3{1+w_i}}\phi_0^{3(1+w_i)h_f}\mathcal{A}_2^{-\frac{3(1+w_i)h_f}{2}}R^{\frac{3(1+w_i)h_f}{2}}
\end{align}
Keeping the dominant term in the limit $R\rightarrow 0$,  we get the final expression for the function $Q(\phi )$,
\begin{equation}\label{r36}
Q(\phi (R))=c_3R^{-\delta }
\end{equation}
Thereby, combining relations (\ref{r29}), (\ref{r36}) and substituting to equation (\ref{r1}), we get the final expression for the reconstructed $F(R)$ gravity in the small $R$ limit,
\begin{equation}\label{r37}
F(R)\simeq c_1-c_3\frac{1}{R^{\delta }}
\end{equation}
Recalling the value of $\delta$ from (\ref{r30}), which is $\delta =1$, the most dominant term for the $F(R)$ gravity is the second one, so finally, the small $R$ reconstructed $F(R)$ gravity reads,
\begin{equation}\label{r38}
F(R)\simeq -c_3\frac{1}{R}
\end{equation} 
Having the early time and late time behavior of the reconstructed $F(R)$ gravity at hand, given in relations (\ref{r38}) and (\ref{r24c}) we have a particularly interesting behavior of the reconstructed $F(R)$ gravity. Particularly, and with regards to the late time $F(R)$ gravity, the late time acceleration is given by a negative power function of the Ricci scalar, with these $F(R)$ gravities being well known to describe late time acceleration (see for example \cite{reviews1,reviews2,importantpapers3}). So the overall behavior of the reconstructed modified gravity, taking large and small curvature regimes simultaneously into account, results to the following $F(R)$ function,
\begin{equation}\label{r39}
F(R)\simeq R+\alpha R^2-c_3\frac{1}{R}
\end{equation}
with the parameter $\alpha $ given in (\ref{r24b}). Therefore, the reconstructed function that generates the Hubble parameter corresponding to the matter bounce holonomy corrected LQC, can actually describe inflation and late-time acceleration in the Jordan frame. However we need to discuss an important issue at this point being related to the small-curvature behavior of the $F(R)$ gravity. It is known that LQC recovers the standard Einstein-Hilbert gravity in the small curvature regime, but in the case we studied in this paper, the small curvature $F(R)$ gravity behaves as $1/R$, so this is a rather disturbing feature of the theory. However, we need to note that the LQC theory is based on adding holonomy corrections to the standard Einstein-Hilbert gravity, therefore in the limit $\rho_c\rightarrow \infty$, the standard Einstein-Hilbert gravity is recovered. For details on this see \cite{LQC7sing,defnew} and section 3 of the recent work \cite{harob}. In addition, notice that in the small curvature limit of the LQC approach, the standard FRW equation of the Einstein-Hilbert gravity is recovered, by taking the limit $\rho_c\rightarrow \infty$ in Eq. (\ref{holcor1}).

In our case however, the pure $F(R)$ gravity limit that reproduces the large $t$ (small curvature) limit of the matter bounce scenario, is not the Einstein-Hilbert gravity but an $1/R$ gravity. This is certainly a difference between the two approaches that needs to be pointed out, but note that in the context of $F(R)$ gravity, many cosmological scenarios that remain ``exotic'' for the standard Einstein-Hilbert gravity, like the cosmological bounces, can be produced by certain $F(R)$ gravity. In addition, we assumed a total absence of matter fluids, so this might be also the reason for producing such a late-time behavior of the pure $F(R)$ gravity. In any case, this is an important issue and we needed to point this out in order to clarify this vague point. Perhaps there exists a deeper mathematical reasoning behind this difference, that should be checked extensively. This task however is beyond the scopes of this article and we hope to address in the future. 

It is interesting to go through an Einstein frame analysis of the implications that such an $F(R)$ gravity has on slow-roll inflation parameters, and try to make contact with recent observations corresponding to this cosmological era, so we provide here a brief account on this. Note that in our case, the parameter $\alpha $ is a free parameter, due to the dependence of $\alpha $ on $c_1$, so it can be adjusted so that the spectral index can agree to some extend with the experimental data. For details the reader is referred to \cite{reviews1,reviews2,sergeistarobinsky}. But before going to the inflation study it worths discussing an issue with regards to Jordan and Einstein frame metrics.

\subsection{An Brief Inflation Study and a Quick Overview of Holonomy Corrected $R^2$ Gravity}

In order to discuss inflation in the Einstein frame, in principle, the metric describing inflation in the Einstein frame has to be a de-Sitter, or quasi de-Sitter, or at least a metric that produces acceleration. But in the reconstruction technique we required a flat FRW metric which under the conformal transformation in the Einstein frame, will produce a metric in the Einstein frame which has to be investigated if it is of de-Sitter type. It is rather difficult in general to achieve this. So we start off with the $R^2$ gravity in the Jordan frame, with the geometry in that frame being described by a general solution to the Einstein frame, corresponding to the $R^2$ gravity. It's conformally related counter-part is described by a flat FRW metric of de-Sitter or quasi de-Sitter type. 

Thinking in this way, by means of a conformal transformation, we  express the Jordan frame quantities in terms of the Einstein frame ones. The analysis we shall adopt can be found in \cite{reviews1,reviews2} and also \cite{sergeistarobinsky}. Following very well known procedures, we quote here the Einstein frame quantities, corresponding to the $F(R)$ function (\ref{r24c}). The scalar potential $V(\sigma )$ is equal to,
\begin{equation}\label{dpanve}
V(\sigma )=\frac{\left(-1+e^{\sqrt{\frac{2}{3}} \sqrt{k^2} \sigma }\right)^2}{8 k^2 \alpha }
\end{equation}
In addition, the slow-roll coupled differential equations, 
\begin{equation}\label{slowrolldiff}
\frac{3H^2}{k^2}\simeq V(\sigma ),{\,}{\,}{\,}3H\dot{\sigma}\simeq -V'(\sigma )
\end{equation}
for the potential (\ref{dpanve}) have to be solved explicitly and the solution will reveal if acceleration occurs. In order a quasi-de-Sitter solution is realized, we must require that the fraction $k^2/\alpha^2$, for which we introduce new notation, and set it equal to $\gamma $, satisfies the following condition,
\begin{equation}\label{slowroollnew}
\gamma\sim M^2,{\,}{\,}{\,}M\ll M_p
\end{equation} 
with $M_p$ the Planck mass. Slow roll inflation can be easily achieved, since $\alpha $ contains the free parameter $c_1$, which can be chosen appropriately (recall the value of $\alpha$ from relation (\ref{r24b})).  Then, during the inflation era (that is when $\sigma\rightarrow -\infty$), the slow roll differential equations (\ref{slowrolldiff}) become,
\begin{equation}\label{slowrolldiff1}
H^2\simeq \frac{\gamma}{12},{\,}{\,}{\,}3H\dot{\sigma}\simeq -\left ( \frac{\gamma }{\sqrt{6k^2}}\right )e^{\sqrt{2k^2/3}\sigma}
\end{equation}
yielding the following solution,
\begin{equation}\label{slowrollsolution}
\sigma (t)\simeq -\sqrt{\frac{3}{2k^2 }}\ln \left ( \sqrt{\frac{2\gamma }{27}}(t_0-t)\right )
\end{equation}
It can be explicitly shown that acceleration indeed occurs for this solution, see \cite{sergeistarobinsky}. Having solution (\ref{slowrollsolution}) at hand, we can easily compute the slow roll parameters for inflation in a straightforward way. The inflation ends at approximately $\sigma_e =-0.17\sqrt{3/(2k^2)}$ and the e-fold number is,
\begin{equation}\label{efoldn}
N\simeq \sqrt{\frac{\gamma }{24}}t_0
\end{equation}
Therefore, the slow roll parameters during inflation are equal to,
\begin{equation}\label{slowrollparmduringinfl}
n_s\simeq 1-\frac{2}{N},{\,}{\,}{\,}r\simeq \frac{12}{N^2}
\end{equation}
which means that for approximately sixty e-folds we have $n_s\simeq 0.967 $ and $r\simeq 0.003$, which are not very far from the actual experimental values. Of course, notice that due to the freedom we have in choosing the parameter $\alpha$, offers the possibility to further restrain the final values of the slow roll parameters. Before closing, it worths mentioning that there is a way to eliminate singularities and ensure consistency with Planck data, even for complicated analytic forms of $F(R)$ gravity. This method relies in adding holonomy corrections directly to $F(R)$ gravity. For the $R^2$ case, this was explicitly done in \cite{mbouncersquarefr} and it worths outlining the basic features of this method. For details see \cite{mbouncersquarefr,mbounce5,holonomyfr}.

Introducing the holonomy corrections, the holonomy corrected Einstein frame FRW equation is,
\begin{equation}\label{einframeholcorr}
\tilde{H}^2=\frac{1}{3}\tilde{\rho}\left ( 1-\frac{\tilde{\rho}}{\tilde{\rho_c}}\right )
\end{equation}
with $\tilde{\rho}_c$ denoting the Einstein frame critical density. Equation (\ref{einframeholcorr}) is an ellipse in the $(\tilde{H},\tilde{\rho })$ plane, in which ellipse, the universe's dynamics is depicted. Actually the universe moves clockwise from the contracting phase to the expanding phase, beginning and ending at the critical point $(0,0)$ and bouncing off only once at $(0,\tilde{\rho })$. For the Holonomy corrected $R^2$ gravity given in relation (\ref{r24c}), the equation describing the scalar field evolution in the Einstein frame is given by,
\begin{align}\label{psievol}
& \ddot{\psi}\psi+3\tilde{H}\dot{\psi }\psi+\frac{1}{\alpha }\Big{(}\tilde{\psi }-1 \Big{)}
\end{align}
The Einstein frame energy density $\tilde{\rho}$ is,
\begin{equation}\label{energydensityeinframe}
\tilde{\rho}=\frac{3\ddot{\psi}^2}{4\psi^2}+\frac{1}{8\alpha \psi^2}\left ( \tilde{\psi}-1\right )^2
\end{equation}
Recalling that the Hubble parameter $\tilde{H}$ is related to the energy-density $\tilde{\rho}$ by the holonomy corrected FRW equation (\ref{einframeholcorr}), the Hubble parameter vanishes at the point $(\psi,\dot{\psi})=(1,0)$ and also at the curve $\tilde{\rho}=\tilde{\rho}_c$. Hence, the universe undergoes contraction and acceleration, with the dynamics being critically determined by the critical point $(\psi,\dot{\psi})=(1,0)$ and the curve $\tilde{\rho}=\tilde{\rho}_c$. Note that in our case, the parameter $\alpha $ may vary, and consequently the final dynamics depend crucially on the $\alpha $ parameter, because this determines the curve $\tilde{\rho}=\tilde{\rho}_c$. Notice that in the context of holonomy corrected $F(R)$ gravity, the singularities of the $F(R)$ theory are eliminated \cite{mbouncersquarefr}.

\section{Bounce Scale Factor in the Small and Large $t$ Limits and Discussion on the Validity of the Reconstruction Method}

Having reconstructed the $F(R)$ gravities that reproduce the Hubble parameter of the matter bounce scenario, it worths discussing how the scale factors behave in the small and large cosmic time limits. Recall that in order to apply the reconstruction technique, we made a critical assumption for the scale factor, that is, we assumed that it takes the form given in relation (\ref{r4}). Of course this exponential form of the scale factor is different from the form of the scale factor characteristic to the matter bounce cosmology, given in relation (\ref{holcorrLQCsol}). Obviously in the large and small cosmic time limits, there has to be some overlap, at least to some extent. This is the core subject of this section, along with some general considerations regarding the validity of the reconstruction technique we used. We start off with the investigation of the validity of the technique we employed. One of the crucial assumptions we made is that the function $h(t)$ (\ref{r7}) is slowly varying and consequently, the differential equation (\ref{r5}) is simplified to a great extent. Due to $h(t)$ being a slowly varying function, we ignored the derivatives of this function. Let us see the behavior of the first derivative $h'(t)$ as a function of $t$, and how this function responses to changes of the $q$ parameter. Recall that $q=\frac{3}{4}\rho_c$ and the critical density $\rho_c$ may vary from model to model. For example in \cite{mbounce4} it was taken equal to $\rho_c\sim \frac{2\sqrt{3}}{\gamma^3}$, with $\gamma$ the Barbero-Immirzi parameter, while in \cite{mbounce11}, it was taken equal to $\rho_c\sim 10^{-9}\rho_{pl}$, with $\rho_{pl}=64\pi^2$ \cite{mbounce4}. Hence, in the former case, $\rho_c\simeq 258.51$ while in the latter case $\rho_c\simeq 631.5\times 10^{-9}$, so it may vary from a quite small number to a number of the order $\sim 10^2$. Let us see how the derivative $h'(t)$ behaves as a function of the parameter $q$. 
\begin{figure}[h]
\centering
\includegraphics[width=15pc]{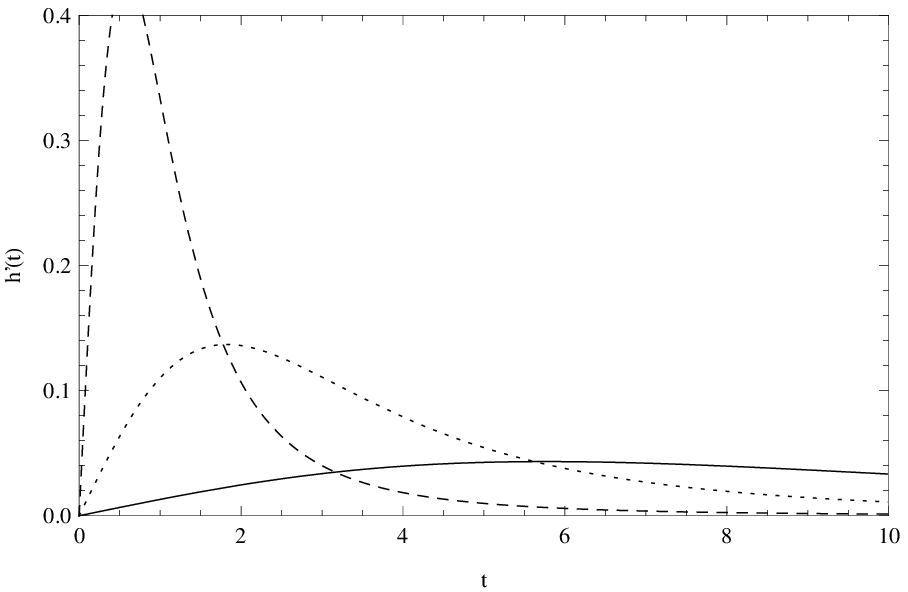}
\includegraphics[width=15pc]{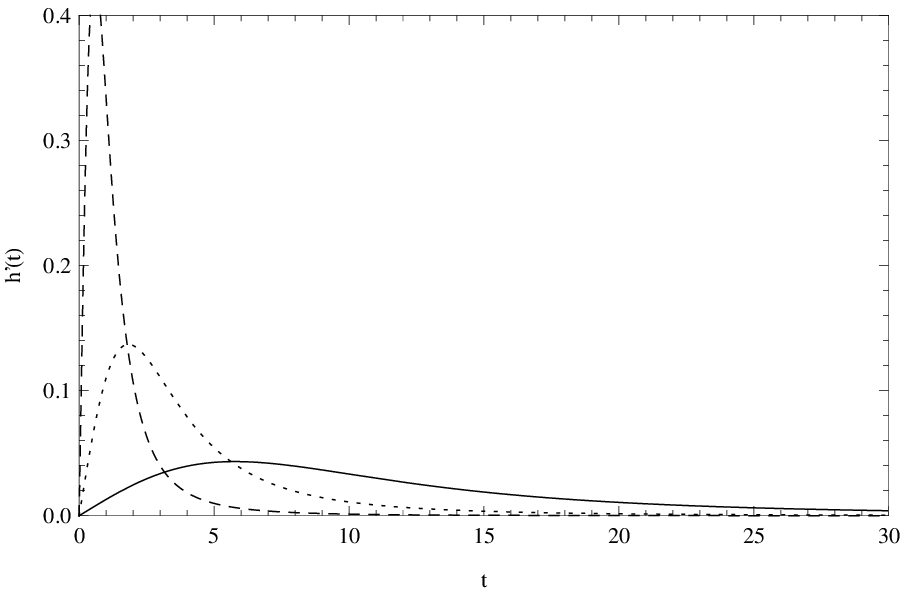}
\caption{Plots of $h'(t)$
as a function of $t$. The dashed, dotted, solid lines refer to the values of $q$, with $q=1$, $q=0.1$ and $q=0.001$ respectively}\label{fig:elizaldeplots1}
\end{figure}
In Fig.1 we presented the plots of the function $h'(t)$ as a function of the cosmic time, for $q=100$, $q=0.1$ and $q=0.001$. In the left figure we can see the behavior of the function for $t\sim O(1)$ while on the right figure we can see the behavior for larger cosmic time values. As we can see from the plots, there is a range of $t$ values for which the function $h'(t)$ takes values which are not negligible, as $q$ grows larger (see the dashed curve corresponding to $q=100$). However when $q$ takes smaller values, the function $h'(t)$ is negligible. 

However, even if the value of $q$ is not small, for small or large cosmic times (and by small and large in this case we mean early cosmic times, when inflation occurs $t\sim 10^{-35}$, and late times which correspond to the present epoch) the derivative is also negligible. So our approximation is valid for small times $t\ll 1$ and large times $t\gg1$. In order to further support this result, let us numerically check the values of $h'(t)$ for various values of $q$ and $t$. In table 1 we have presented the values of $h'(t)$ for various $t$ and as we can see, the approximation $h'(t)\sim 0$ is valid for small times $t\ll 1$ and large times $t\gg1$.
\begin{center} 
    \begin{tabular}{ | p{3cm} | p{3cm}  | p{3cm}  |p{3cm}  |}
    \hline
    $q$  & $t\sim 10^{-35}$ & $t\sim 10^{5}$ & $t\sim 1$ \\ \hline
    $q=10^{-5}$ & $h'(t)\simeq 1.3310^{-40}$ & $h'(t)\simeq 1.3310^{-10}$ & $h'(t)\simeq 0.00001$
     \\ \hline
    $q=0.01$ & $h'(t)\simeq 1.3310^{-38}$ & $h'(t)\simeq 1.3310^{-10}$  & $h'(t)\simeq 0.0013$
    \\ \hline
    $q=1$ & $h'(t)\simeq 1.3310^{-35}$ & $h'(t)\simeq1.3310^{-15} $ & $h'(t)\simeq 0.33$
    \\ \hline
    $q=100$ & $h'(t)\simeq 1.3310^{-33}$ & $h'(t)\simeq 1.3310^{-17}$  & $h'(t)\simeq 0.013$
\\ \hline
    
    \end{tabular}
    \\
    \bigskip 
    Table 1: Values of $h'(t)$ as a function of $t$.
\end{center}
The same arguments apply for the higher derivatives as we can see in Fig. 2, where we have plotted the behavior of the higher derivatives $h''(t)$, $h'''(t)$, $h^{(4)}$, as a function of $t$ for $q=0.1$. As it can be seen, the fourth derivative disappears from the plot. The same applies for even higher derivatives.
\begin{figure}[h]
\centering
\includegraphics[width=15pc]{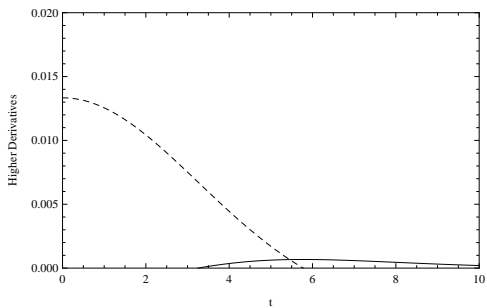}
\caption{Plots of $h''(t)$, $h'''(t)$, $h^{(4)}$
as a function of $t$ for $q=0.1$. The dashed, dotted, solid lines refer to $h''(t)$, $h'''(t)$, $h^{(4)}$ respectively}\label{fig:elizaldeplots1}
\end{figure}
Now let us come to the issue of the scale factor behavior within our approximation and it's comparison to the matter bounce scale factor. It is to be understood of course that the reconstruction method we use is an approximation method, so a complete overlap of the scale factor (\ref{r4}) with the matter bounce one (\ref{holcorrLQCsol}) is rather unlikely. This can easily be seen by substituting $g(\phi)$ from relation (\ref{r11}) to (\ref{r4}), so that the scale factor of the approximation method we use is,
\begin{equation}\label{r40}
a(t)=a_0\left ( \frac{t}{t_0}\right )^{\frac{h_fq_ct^2}{1+q_ct^2}}
\end{equation}
which is a completely different functional form in reference to (\ref{r4}). However, it expected that overlap should be achieved in the small $t$ and large $t$ limits, and this is what we demonstrate here explicitly. We start off with the large $t$ limit first, in which case the matter bounce scale factor is approximately equal to,
\begin{equation}\label{r40a}
a(t)\simeq q^{1/3}t^{2/3}
\end{equation}
with $q=3\rho_c/4$. In addition, the reconstruction scale factor (\ref{r40}) in the large $t$ limit reads,
\begin{equation}\label{r40b}
a(t)\simeq \frac{a_0}{t_0^{h_f}}t^{h_f}
\end{equation}
with $h_f=2/3$. Now if we require,
\begin{equation}\label{r40c}
\frac{a_0}{t_0^{h_f}}=q^{1/3}
\end{equation}
we have exact coincidence of the two scale factors. Now Let us focus on the small $t$ limit, in which case the matter bounce scale factor approximately is,
\begin{equation}\label{r41}
a(t)\simeq 1+\frac{qt^2}{3}
\end{equation}
with $q=3\rho_c/4$. On the other hand, the scale factor within the context of the reconstruction method in the small $t$ limit reads,
\begin{equation}\label{r42}
a(t)=a_0+a_0h_fq\ln \left (\frac{t}{t_0} \right )t^2
\end{equation}
which is at zero order in $t$ identical to (\ref{r41}) if $a_0=1$. Now let us see how the function of the above relation (\ref{r42}) behaves at small $t$ and for various values of $q$ in comparison to the behavior of (\ref{r41}), bearing in mind the constraint (\ref{r40c}). In Fig.3 we have plotted the functional dependence of relations (\ref{r41}) and (\ref{r42}) as a function of $t$, for $q=10^{-5}$ (left) and for $q=1$ (right).  
\begin{figure}[h]
\centering
\includegraphics[width=15pc]{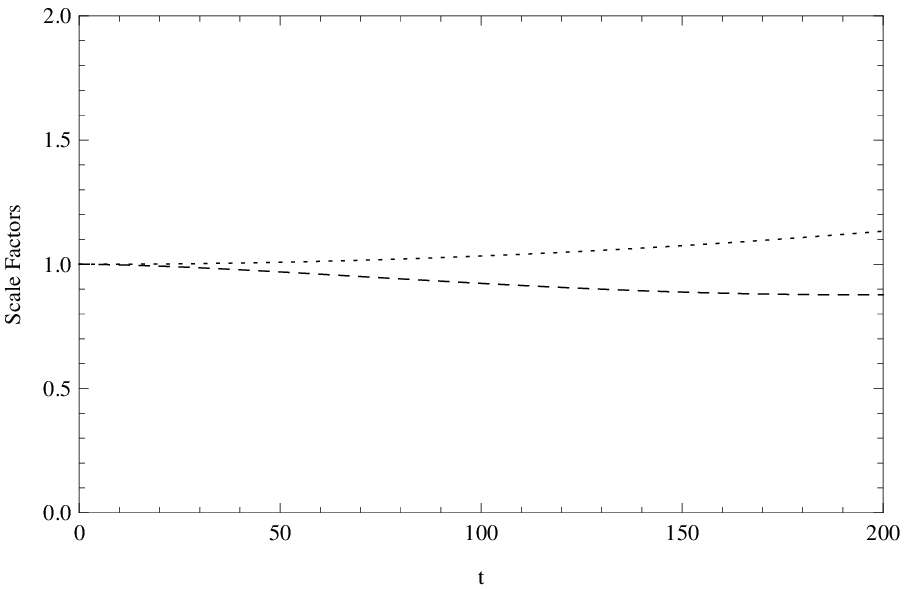}
\includegraphics[width=15pc]{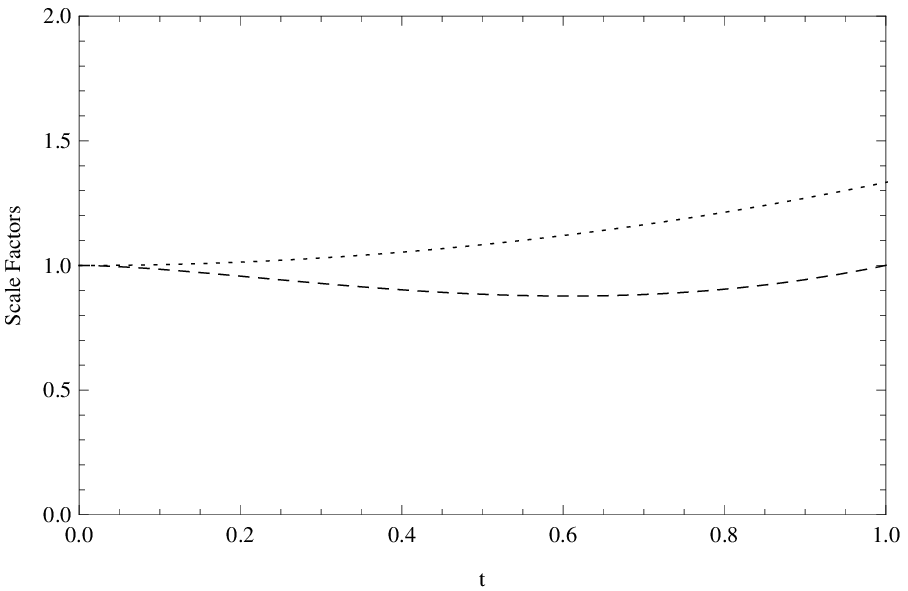}
\caption{Plots of the scale factors $a(t)\simeq 1+\frac{qt^2}{3}$ (dotted) and $a(t)=a_0+a_0h_fq\ln \left (\frac{t}{t_0} \right )t^2$ (dashed)
as a function of $t$, for $q=10^{-5}$ (left) and $q=1$ (right)}\label{fig:elizaldeplots1}
\end{figure}
As can easily be seen, the scale factors for small $q$, are identical for a quite large value of the cosmic time $t$, far beyond where the approximation is valid, so absolute concordance is ensured for small $t$, with $t \ll 1$. The same applies for the $q=1$ case, with the difference that the scale factors are identical only for $t\ll 1$, which in any case is the case which interests us. In table 2 we present the values of the scale factors for $q=10^{-5}$ for various values of time $t$ and the conclusion is that for $t\ll 1$ our approximation is valid and the scale factors of the bounce matter scenario and of our approximate method are equivalent. 
 \begin{center} 
    \begin{tabular}{ | p{5cm} | p{3cm}  | p{3cm}  |p{3cm}  |}
    \hline
    Scale Factors for $q=10^{-5}$   & $t\sim 10^{-35}$ & $t\sim 10^{-5}$ & $t\sim 1$ \\ \hline
    $a(t)=a_0+a_0h_fq\ln \left (\frac{t}{t_0} \right )t^2$ & 1 & 0.999999999999988  & 0.999961623
     \\ \hline
    $a(t)= 1+\frac{qt^2}{3}$ & 1 & 1  & 1.00000333
    \\ \hline 
    \end{tabular}
    \\
    \bigskip 
    Table 2: Values of Scale Factors for $q=10^{-5}$.
\end{center}
It remains to formally explain why we do not have explicit functional coincidence between the two scale factors. It is to be understood that we are working in an approximate method, but in a later section, we shall formally address this issue from a mathematical point of view.

\subsection{A Brief Comparison with Other Bounce Solutions}

In this small section we shall try to reveal a link between the reconstructed $F(R)$ gravity we presented in this section in the small $t$ limit, namely equation (\ref{r24c}), to the bounce solutions presented in reference \cite{sergeibounce}. Recall that the $F(R)$ gravity solution we found (\ref{r24c}) correspond to the small $t$ limit of the matter bounce scenario in LQC scale factor, given in equation (\ref{holcorrLQCsol}). Note that this limit corresponds to the inflationary period of the universe. It's worth describing the bounce solutions given in reference \cite{sergeibounce}, in order to be informative as possible. As was explicitly shown in \cite{sergeibounce}, using the reconstruction method \cite{importantpapers12}, the bounce solution with scale factor $a(t)$ given by the following expression,
\begin{equation}\label{bouncesolsergeijcap}
a(t)\simeq e^{\alpha_1 t^2}
\end{equation}
can be produced by the following $F(R)$ gravity,
\begin{equation}\label{prodfrgravity}
F(R)=\frac{1}{\alpha_1}R^2-72R+144 \alpha_1
\end{equation}
It was assumed that the universe is described by a flat FRW metric of the form (\ref{metricformfrwhjkh}). The solution (\ref{bouncesolsergeijcap}) actually describes a bounce around $t=0$, in which case for $t<0$, the Hubble parameter is negative, that is $H<0$ (contracting phase), and for $t>0$, the Hubble parameter is positive, $H>0$ (expanding phase). The bounce solution (\ref{bouncesolsergeijcap}) can be related to the matter bounce solution we presented in this section, given in (\ref{r41}) and in order to see this, we expand the scale factor (\ref{bouncesolsergeijcap}) around $t=0$ and we get, keeping the lowest order in  $t$,
\begin{equation}\label{taylorexp}
a(t)\simeq 1+\alpha_1 t^2
\end{equation}
Hence, for $\alpha_1=\frac{q}{3}$, the two scale factors are identical. So the small $t$ limit of the bounce solution (\ref{bouncesolsergeijcap}), maybe be produced by the reconstructed $F(R)$ gravity (\ref{r24c}), which for small $t$ generates a scale factor that behaves as the matter bounce scale factor in the same limit. Thereby, there is an indirect correlation between the $F(R)$ gravity (\ref{prodfrgravity}) with the one given by relation (\ref{r24c}). However, the gravity (\ref{prodfrgravity}) cannot describe Starobinsky inflation, as was argued in \cite{sergeibounce}. In the present case however, the reconstructed function can generate Starobinsky inflation, because of the freedom we have in the choice of the parameter $\alpha$, which depends on the free parameter $c_1$, see relation (\ref{r24b}).

\section{Discussion}

In this section we shall attempt to explain from a mathematical perspective how the reconstruction method with an auxiliary field works, in order to understand how the validity of the method is ensured and also how the method succeeds in reconstructing the matter bounce cosmological solutions, at least in the small and large cosmic time $t$ limits. The starting point of our analysis shall be action (\ref{neweqn123}) and particularly the Lagrangian density entering this action. For the mathematical concepts to be used in this section, the reader is referred to \cite{shad}. The total spacetime is a non-trivial fibre bundle $Y\rightarrow X$, with $Y\equiv R^4$ and $X\equiv R$. The variation of the action with respect to the metric, with the metric being a flat FRW metric, yields the Euler-Lagrange equations (\ref{neweqn123}), and by solving these we obtain the scale factor $a(t)$ which determines the metric that maximizes the action. In order to put the above in a mathematical context, we shall make use of the jet bundle of the total fibre bundle. Since the Euler-Lagrange equations  for general relativity result from second order Lagrangian densities, these lead to second order Euler-Lagrange equations \cite{shad}. The solutions of these equations of motion, are local sections (from now on the sections are considered to be local sections only) of the kernel of second order Euler-Lagrange operators, corresponding to the specific Euler-Lagrange equations under study. To get into details, these equations of motion in our case constitute a system of second order differential equations which is formally a closed subbundle of the jet bundle $\mathcal{J}^2Y\rightarrow X$. We denote the closed subbundle as $\mathcal{G}$ for the purposes of this article. Before continuing our discussion, let us briefly recall the essentials of jet bundles, to render the presentation self-contained. A $k$-order jet manifold of a total fibre bundle $Y\rightarrow X$, which is denoted as $\mathcal{J}^2Y$, provides the equivalence classes of the (local) sections $s$ of $Y$, which are identified by the Taylor series of these sections around any $x$ $\in$ $X$. Practically, the jet manifold is described by the coordinates $(x^{\lambda}, y^i_{\lambda_1},y^i_{\lambda_1,\lambda_2},...,y^i_{\lambda_1,...,\lambda_k})$, with $y^i_{\lambda_1,...,\lambda_l}=\partial_{\lambda_1}\partial_{\lambda_{2}}...\partial_{\lambda_l}s^i(x)$, with $0\leq l\leq k$. Thus, there is an surjective (onto) mapping between the sections $s^i(x)$ and the sections $\mathcal{J}^ks$ of the jet bundle $\mathcal{J}^2Y\rightarrow X$, such that the following relations holds true,
\begin{equation}\label{bun1}
y^i_{\lambda_1,...,\lambda_k}\circ\mathcal{J}^ks=\partial_{\lambda_1}\partial_{\lambda_{2}}...\partial_{\lambda_l}s^i(x)
\end{equation} 
with $0\leq l\leq k$. Returning to the case at hand, the solution to the system of second order differential equations is a local section $s$ of the fibre bundle $Y\rightarrow X$, such that its jet prolongation $\mathcal{J}^2s$, lives in the closed subbundle $\mathcal{G}$ of the jet bundle $\mathcal{J}^2Y$. These differential equations are associated with differential operators taking their values in a vector bundle $E\rightarrow X$, coordinated by $(x^{\lambda},v^A)$. In the case at hand, the differential equations are second order Euler-Lagrange equations and the vector bundle in which the second order Euler-Lagrange operators take their values is locally $E\equiv T^*Y \wedge T^*X$, with $T^*Y,T^*X $ the dual tangent spaces of the spaces $Y,X$ (practically the vector bundle $E$ is the space of one-forms on $Y,X$, differentials so to speak). The second order Euler-Lagrange operator is defined to be a fibred morphism which locally acts as,
\begin{align}\label{fm}
& \mathcal{E}:\mathcal{J}^2Y\rightarrow E \\ \notag &
v^A\circ \mathcal{E}=\mathcal{E}^A(x^{\lambda}, y^i_{\lambda_1},y^i_{\lambda_1,\lambda_2},...,y^i_{\lambda_1,...,\lambda_k})
\end{align}
Practically speaking, the operator sends each section $s(x)$ of the fibre bundle $Y\rightarrow X$, onto the section $\mathcal{E}\circ \mathcal{J}^2Ys(x)$ of the vector bundle $E\rightarrow X$, that is, $(\mathcal{E}\circ \mathcal{J}^2Ys)^A =\mathcal{E}^A(x^{\lambda}, s^i(x),\partial_{\lambda_1}s^i(x),\partial_{\lambda_2}s^i(x))$. Assume that $O(x)\subset \mathcal{E}(\mathcal{J}^2Y)$ is the set containing the zero sections of the vector bundle $E\rightarrow X$. Then, the Kernel of the Euler-Lagrange operator $\mathcal{E}$ is,
\begin{equation}\label{ker}
\mathrm{ker}\mathcal{E}\subset \mathcal{E}^{-1}(O(x))\subset \mathcal{J}^2Y
\end{equation}
If $\mathrm{ker}\mathcal{E}$ is a closed subbundle of $\mathcal{J}^2Y$, the second order Euler-Lagrange equations are formally defined to be,
\begin{equation}\label{kerrdiffeqns}
\mathcal{E}\circ \mathcal{J}^2s=0
\end{equation}
The method we used in this paper works in two possible ways, which we describe in detail now. So in the first case, practically by applying our method it is equivalent to choosing a subbundle of $\mathrm{ker}\mathcal{E}$, which we denote $A'$ (practically this would be the exponential form of the scale factor), so that,
\begin{equation}\label{ker1}
A'\subset \mathrm{ker}\mathcal{E}\subset \mathcal{J}^2Y
\end{equation}
This holds true since we assume a specific form of solutions (exponential) for the sections of the kernel of the Euler-Lagrange operator. So this actually restricts the differential operator further to a subbundle of the vector bundle $E$, which means that the new Euler-Lagrange operator is a restriction of the initial morphism which defines the vector bundle $E$, to the subbundle $E'$ of $E$, that is,
\begin{align}\label{fm1}
& E'\subset E \\ \notag &
\mathcal{E}':\mathcal{J}^2\rightarrow E'=\mathcal{E}(\mathcal{J}^2Y)\mid_{E'}
\end{align}
Practically when we use the reconstruction method, we actually try to find these second order Euler-Lagrange operators $\mathcal{E}'$. The terminology restriction is not accidental, since this new morphism of the new vector bundle $E'$, is a restriction of the initial vector bundle $E$, owing to the slow varying function $h(t)$ approximation, according to which we disregarded the higher derivatives of $h(t)$ in the Euler-Lagrange differential equations. Now the key point is that the reconstruction technique leads to the same Hubble parameter with the matter bounce scenario, a fact that hold true if we disregard $h'(t),h''(t)...$, etc. The reconstruction technique leads to a Hubble parameter $H(t)$ which is given by relation (\ref{r11a}). Therefore, there exist two morphisms, which we denote $H_1$ and $H_2$ such that,
\begin{equation}\label{morphubble}
H_1: s'\rightarrow H_1(s')\mid_{H_2(s)},{\,}{\,}{\,}H_2:s\rightarrow H_2(s)
\end{equation}  
where $H_1(s')$ and $H_2(s)$ are the reconstruction method and matter bounce Hubble parameters respectively and $s,s'$ are the scale factors for the reconstruction method and matter bounce scenario respectively (local sections of the kernel of the vector bundle $\mathcal{E}'\subset \mathcal{J}^2Y$). Notice the form of the morphism $H_1$, which implies that in general $H_2(s)\subset H_1(s')$ and consequently, by disregarding all the higher derivatives of $h(t)$, the following equivalence relation holds true,
\begin{equation}\label{equiv}
H_1(s')\mid_{H_2(s)}\equiv H_2(s)
\end{equation}
So practically the following holds true,
\begin{equation}\label{holds}
H_1: \mathcal{J}Y\rightarrow H_1(s')\mid_{H_2(s)}
\end{equation}
The fact that there is overlap between the matter bounce scale factor and the reconstruction method scale factor in the large and small $t$ limits, is due to the equality of the corresponding scale factors, or more formally stated because of the morphism equivalence (\ref{equiv}). We have to note that in general however, $H_1(s')\subset H_2(s)$ because the Hubble parameter corresponding to the reconstruction method gives approximately ($h'(t)\sim 0$) the matter bounce Hubble parameter. 

The second case arises because the assumed relation $\mathrm{ker}\mathcal{E}'\subset \mathrm{ker}\mathcal{E} $, is not always true. In such a case, the generalized version of this is, $\mathrm{ker}\mathcal{E}'\bigcap \mathrm{ker}\mathcal{E} \neq \varnothing$. In this case, the operator $\mathcal{E}'$ is not simply a restriction on the original subbundle $\mathrm{ker}\mathcal{E}$. Of course, the existence of an exponential form for the scale factor ensures that $\mathrm{ker}\mathcal{E}'$ is a closed subbundle of the jet bundle $\mathcal{J}^2Y$. Apart from this, the rest of the argument is pretty much the same, with the equivalence (\ref{equiv}) holding true in this case too and playing an important role in supporting the argument. We hope to address these interesting mathematical issues in more details in a future work.

\section*{Concluding Remarks}

In this paper we applied the reconstruction technique with an auxiliary field, in order to find which $F(R)$ gravity can reproduce the cosmological solutions of the matter bounce scenario. Particularly, owing to the functional form of the Hubble parameter corresponding to the matter bounce scenario, the reconstruction technique we used leads after some simplifications, to the Hubble parameter of the matter bounce scenario. Then by studying the problem at hand in the small and large cosmic time limits, we were able to find which $F(R)$ gravity gives this Hubble parameter in each case. Particularly, in the small $t$ limit, which corresponds to large values of the scalar curvature, the reconstructed $F(R)$ gravity is $F(R)\simeq R+\alpha R^2$, a gravity that is known to reproduce Starobinsky inflation. Practically, the positive power of the scalar curvature is known to produce inflation at the early stages of the universe's evolution \cite{importantpapers3}. Interestingly enough, in the large $t$ limit, which corresponds to small scalar curvatures, the $F(R)$ gravity is of the form $F(R)\simeq R-c_3\frac{1}{R}$, which is known to produce late-time acceleration \cite{importantpapers3}. The reconstruction technique leads to a scale factor $a(t)$ which, at the large and small cosmic time limits, coincides almost exactly with the matter bounce scenario scale factor in the corresponding limits.

So we are confronted with the appealing physical picture of having a Jordan frame $F(R)$ gravity, which in the small $t$ and large $t$ limit imitates the cosmological behavior of the matter bounce scenario cosmological solutions. Interestingly enough, this $F(R)$ gravity at large curvatures produces Starobinsky inflation and also late-time acceleration in the Jordan frame. Also, regarding the inflation, it can be adjusted to produce Einstein frame Starobinsky inflation agreeing with the Planck observational data, owing to the freedom we have in choosing the parameter $\alpha$. Of course the full $F(R)$ gravity is much more complicated than these limiting gravities we found, so it is needed to construct much more complex theoretical frameworks, so that we can have agreement with observational data. Such a framework, the holonomy corrected $F(R)$ gravity is done in the literature \cite{mbouncersquarefr} and we briefly mentioned in the text the most important features of this theory. 

Since the method we used is an approximation and is based on specific assumptions, we explicitly studied to which extent the approximations we did are valid. We demonstrated that the approximation is valid to a great extent and for a large range of the parameter values. A mathematical explanation of how the reconstruction method works was presented in section 3.

In addition, since the method is an approximate method based on the fact that $\phi=t$, one has to check explicitly the stability of the solutions we found and also check the local implications of these gravities to astrophysical objects and local gravity constraints. We plan to address these issues in a future work.

Before closing, let us note that the matter bounce scenario is an appealing alternative to Big Bang and it is an experimental challenge to verify it's seeds in observations. Apart from the Planck and BICEP experiments, there exist alternative methods coming from direct observations of dark matter \cite{kinezosvergados}, that may reveal the bounce behavior of the universe. It's certainly worth trying to make contact with these indirect methods of observing the universe's evolution in particle physics systems. For indirect methods of observing dark matter, see for example \cite{oikonomouvergados} and references therein.

\section*{APPENDIX A}

In this appendix we provide the full details for the polynomial coefficients $\alpha_i,\beta_j$, with $i=0,..3$ and $j=0,...5$, appearing in relation (\ref{r15}) Recall for convenience that the $\alpha_i$'s are coefficients of the polynomial:
\begin{equation}\label{a1}
\rho (R)=\alpha_0+\alpha_1R+\alpha_2R^2+\alpha_3R^2
\end{equation} 
while the $\beta_j$'s are coefficients of the polynomial,
\begin{equation}\label{a2}
P(R)= \beta_2R^2+\beta_3 R^3+\beta_4 R^4+\beta_5R^5
\end{equation}
and these are given below,
\begin{align}\label{a3}
& \alpha_0=432 q^6-216 q^6 h_f+468 q^6 h_f^2-146 q^6 h_f^3+156 q^6 h_f^4-24 q^6 h_f^5+16 q^6 h_f^6 \\ \notag &
\alpha_1=216 q^5-18 q^5 h_f-75 q^5 h_f^2+30 q^5 h_f^3-48 q^5 h_f^4 \\ \notag & 
\alpha_2= 36 q^4+3 q^4 h_f+30 q^4 h_f^2 \\ \notag &
\alpha_3= 2 q^3 \\ \notag &
\beta_2= 45684 q^{10} h_f^2-15228 q^{10} h_f^3+31725 q^{10} h_f^4-5076 q^{10} h_f^5+5076 q^{10} h_f^6 \\ \notag &
\beta_3 = 23004 q^9 h_f^2+594 q^9 h_f^3-12528 q^9 h_f^4-216 q^9 h_f^5 \\ \notag &
\beta_4 =3861 q^8 h_f^2+540 q^8 h_f^3-108 q^8 h_f^4 \\ \notag &
\beta_5 = 216 q^7 h_f^2
\end{align}


\begin{thebibliography}{}






\bibitem{riess} A.G. Riess et al. (High-z Supernova Search Team), Astronom. J. 116, 1009 (1998) [arXiv:astro-ph/9805201]

\bibitem{planck} P.A.R. Ade et al. [arXiv:1302.5082]

\bibitem{bicep} P. A. R. Ade et al. [arXiv:1403.3985], (2014)

\bibitem{reviews1} S. Nojiri, S. D. Odintsov, Int.J.Geom.Meth.Mod.Phys. 11 (2014) 1460006 [arXiv:1306.4426]; Int. J. Geom. Meth. Mod.Phys. 4 (2007) 115 [hep-th/0601213]
 

\bibitem{reviews2} S. Capozziello, V. Faraoni, Beyond Einstein Gravity, Springer, Berlin 2010

\bibitem{reviews3} A. de la Cruz-Dombriz, D. Saez-Gomez, Entropy 14 (2012) 1717 [arXiv:1207.2663]; F. S. N. Lobo, Dark Energy-Current Advances and Ideas, 173-204 (2009) [arXiv:0807.1640]

\bibitem{reviews4} S. Nojiri, S. D. Odintsov,  Phys.Rept. 505 (2011) 59 [arXiv:1011.0544]

\bibitem{reviews5} S. Capozziello, M. De Laurentis, Phys.Rept. 509 (2011) 167 [arXiv:1108.6266]


\bibitem{reviews8} K. Bamba, S. Nojiri, S. D. Odintsov, JCAP 0810 (2008) 045 [arXiv:0807.2575]

\bibitem{reviews9} S. Nojiri, S. D. Odintsov, Phys.Lett. B657 (2007) 238 [arXiv:0707.1941]



\bibitem{importantpapers1} S. Capozziello, S. Nojiri, S.D. Odintsov, A. Troisi, Phys.Lett. B639 (2006) 135 [astro-ph/0604431]; S. Nojiri, S. D. Odintsov, Phys.Rev. D77 (2008) 026007 [arXiv:0710.1738]

\bibitem{importantpapers2} S. Capozziello, V.F. Cardone, S. Carloni, A. Troisi, Int.J.Mod.Phys. D12 (2003) 1969 [astro-ph/0307018]

\bibitem{importantpapers3} S. Nojiri, S. D. Odintsov, Phys.Rev. D74 (2006) 086005 [hep-th/0608008]; A. de la Cruz-Dombriz, A. Dobado, Phys.Rev. D74 (2006) 087501 [gr-qc/0607118] 

\bibitem{importantpapers4} W. Hu, I. Sawicki, Phys.Rev.D76 (2007) 064004 [arXiv:0705.1158] 

\bibitem{importantpapers5} S. M. Carroll, V. Duvvuri, M. Trodden, M. S. Turner, Phys.Rev. D70 (2004) 043528 [astro-ph/0306438]; S. Capozziello, Int.J.Mod.Phys.D11, 483 (2002) [gr-qc/0201033]

\bibitem{importantpapers6} R. Myrzakulov, L. Sebastiani, S. Zerbini, Int.J.Mod.Phys. D22 (2013) 1330017 [arXiv:1302.4646]


\bibitem{importantpapers8} O. Bertolami, R. Rosenfeld, Int.J.Mod.Phys. A23 (2008) 4817 [arXiv:0708.1784]

\bibitem{importantpapers9} A. Capolupo, S. Capozziello, G. Vitiello, Int.J.Mod.Phys. A23 (2008) 4979 [arXiv:0705.0319]

\bibitem{importantpapers10}   P. K.S. Dunsby, E. Elizalde, R. Goswami, S. Odintsov, D. S. Gomez, Phys.Rev. D82 (2010) 023519 [arXiv:1005.2205]

\bibitem{importantpapers11} G. Cognola, E. Elizalde, S. Nojiri, S.D. Odintsov, L. Sebastiani, S. Zerbini, Phys.Rev. D77 (2008) 046009 [arXiv:0712.4017 ]; K. Bamba, Chao-Qiang Geng, Chung-Chi Lee, JCAP 1008 (2010) 021 [arXiv:1005.4574]


\bibitem{importantpapers12}   S. Nojiri, S. D. Odintsov, D. Saez-Gomez, Phys.Lett. B681 (2009) 74 [arXiv:0908.1269]  

\bibitem{importantpapers13} S. Capozziello, V. F. Cardone, A. Troisi, Phys.Rev. D71 (2005) 043503 [astro-ph/0501426]

\bibitem{importantpapers14} J. C.C. de Souza, Valerio Faraoni, Class.Quant.Grav. 24 (2007) 3637 [arXiv:0706.1223]; V. Faraoni, Phys.Rev. D74 (2006) 104017 [astro-ph/0610734];G. J. Olmo, Phys.Rev.Lett. 95 (2005) 261102 [gr-qc/0505101]; G. J. Olmo, Phys.Rev. D75 (2007) 023511 [gr-qc/0612047] 

\bibitem{importantpapers15} S. A. Appleby, R. A. Battye, A. A. Starobinsky, JCAP 1006 (2010) 005 [arXiv:0909.1737]
 

\bibitem{importantpapers17} S. A. Appleby, R. A. Battye, Phys.Lett.B654 (2007) 7 [arXiv:0705.3199]; S. A. Appleby, R. A. Battye, JCAP 0805 (2008) 019 [arXiv:0803.1081]

\bibitem{importantpapers18} A. Silvestri, M. Trodden, Rept. Prog. Phys. 72 (2009) 096901 [arXiv:0904.0024]

\bibitem{importantpapers19} E. Elizalde, E.O. Pozdeeva, S.Yu. Vernov, Phys.Rev. D85 (2012) 044002 [arXiv:1110.5806]

\bibitem{importantpapers20} V. Faraoni,  Phys.Rev. D75 (2007) 067302 [gr-qc/0703044]


\bibitem{sergeinojirimodel} S. Nojiri, S. D. Odintsov, Phys.Rev. D68 (2003) 123512 [hep-th/0307288]

\bibitem{capo} M. Sami, Curr. Sci. 97,887(2009) [arXiv:0904.3445]; Yi-Fu Cai, E. N. Saridakis, M. R. Setare, Jun-Qing Xia, Phys.Rept. 493 (2010) 1 [ arXiv:0909.2776]

 \bibitem{capo1} T. Padmanabhan, Phys.Rept. 380 (2003) 235 [hep-th/0212290]; K. Bamba, S. Capozziello, S. Nojiri, S. D. Odintsov, Astrophys. Space Sci. 342, 155 (2012) [arXiv:1205.3421]

\bibitem{peebles} P.J.E. Peebles, Bharat Ratra, Rev.Mod.Phys. 75 (2003) 559 [astro-ph/0207347]; V. Sahni, AIP Conf.Proc. 782 (2005) 166, J.Phys.Conf.Ser. 31 (2006) 115; M. Li, Xiao-Dong Li, S. Wang, Yi Wang, Commun.Theor.Phys. 56 (2011) 525 [arXiv:1103.5870]; A. Joyce, B. Jain, J. Khoury, M. Trodden [arXiv:1407.0059]

\bibitem{faraonquin} V. Faraoni, Int.J.Mod.Phys. D11 (2002) 471 [astro-ph/0110067]; V.K. Onemli, R.P. Woodard, Class.Quant.Grav. 19 (2002) 4607 [gr-qc/0204065] 

\bibitem{tsujiintjd} A. Gomez-Valent, J. Sola, S. Basilakos, arXiv:1409.7048; S. Basilakos, S. Nesseris, L. Perivolaropoulos, Phys.Rev. D87 (2013) 12, 123529 [arXiv:1302.6051]
 
\bibitem{quintense} Md. Wali Hossain, R. Myrzakulov, M. Sami, E. N. Saridakis, Phys.Rev. D89 (2014) 123513
 
\bibitem{saridakismyrzakulov} Md. Wali Hossain, R. Myrzakulov, M. Sami, Emmanuel N. Saridakis, arXiv:1410.6100 
 

\bibitem{LQC1}A. Ashtekar, P. Singh, Class. Quant. Grav. 28, 213001 (2011) [arXiv:1108.0893 ]

\bibitem{LQC2} A. Ashtekar, Nuovo Cim. B122 (2007) 135 [gr-qc/0702030]


\bibitem{LQC3} A. Corichi, P. Singh, Phys.Rev. D80 (2009) 044024 [arXiv:0905.4949]

\bibitem{LQC4} P. Singh, K. Vandersloot, G.V. Vereshchagin, Phys.Rev. D74 (2006) 043510 [gr-qc/0606032]

\bibitem{LQC5sing} P. Singh, Class.Quant.Grav. 26 (2009) 125005 [arXiv:0901.2750]

\bibitem{LQC6sing} A. Ashtekar, T. Pawlowski, P. Singh, Phys.Rev. D74 (2006) 084003 [gr-qc/0607039]

\bibitem{LQC7sing} M. Bojowald, Class.Quant.Grav. 26 (2009) 075020 [arXiv:0811.4129]


\bibitem{LQC7sing1} M. Sami, P. Singh, Shinji Tsujikawa, Phys.Rev. D74 (2006) 043514


\bibitem{LQC8} E.J. Copeland, D.J. Mulryne, N.J. Nunes, M. Shaeri, Phys.Rev. D77 (2008) 023510 [arXiv:0708.1261]

\bibitem{LQC9} D. Samart, B. Gumjudpai, Phys.Rev. D76 (2007) 043514 [gr-qc/0605113]

\bibitem{LQC10} T. Naskar, J. Ward, Phys.Rev. D76 (2007) 063514 [arXiv:0704.3606 ]



\bibitem{LQC11} T. Cailleteau, J. Mielczarek, A. Barrau, J. Grain, Class.Quant.Grav. 29 (2012) 095010 [arXiv:1111.3535 ]

\bibitem{LQC12} T. Cailleteau, A. Barrau, J. Grain, F. Vidotto, Phys.Rev. D86 (2012) 087301 [arXiv:1206.6736]

\bibitem{LQC13} T. Cailleteau, A. Barrau, J. Grain, F. Vidotto, Phys.Rev. D86 (2012) 087301 [arXiv:1206.6736]



\bibitem{mbounce1} R. H. Brandenberger [arXiv:1206.4196; J. Quintin, Yi-Fu Cai, R. H. Brandenberger, Phys. Rev. D90 (2014) 063507 [arXiv:1406.6049 ]; Yi-Fu Cai, D. A. Easson, R. Brandenberger, JCAP 1208 (2012) 020 [arXiv:1206.2382] ; Yi-Fu Cai, R. Brandenberger, X. Zhang, Phys.Lett. B703 (2011) 25 [arXiv:1105.4286] 

\bibitem{mbounce2} Yi-Fu Cai, R. Brandenberger, X. Zhang, JCAP 1103 (2011) 003 [arXiv:1101.0822]; C. Li, R. H. Brandenberger, Yeuk-Kwan E. Cheung [arXiv:1403.5625; Yi-Fu Cai, E. McDonough, F. Duplessis, R. H. Brandenberger, JCAP 1310 (2013) 024 [arXiv:1305.5259]; R. H. Brandenberger [arXiv:1206.4196  


\bibitem{mbounce3} P. Singh, Class.Quant.Grav. 26 (2009) 125005 [arXiv:0901.2750]


\bibitem{mbounce4} J. Amoros, J. Haro, S. D. Odintsov, Phys.Rev. D87 (2013) 104037 [arXiv:1305.2344]; T. Qiu, X. Gao, E. N. Saridakis, Phys.Rev. D88 (2013) 4, 043525 [arXiv:1303.2372]


\bibitem{mbounce5} J. Haro, Europhys. Lett. 107 (2014) 29001 [arXiv:1403.4529]


\bibitem{mbounce6} Yi-Fu Cai, Shih-Hung Chen, J. B. Dent, S. Dutta, E. N. Saridakis, Class.Quant.Grav. 28 (2011) 215011 [arXiv:1104.4349]


\bibitem{mbouncersquarefr} J. Amoros, J. de Haro, S.D. Odintsov, Phys.Rev. D89 (2014) 104010 [arXiv:1402.3071]


\bibitem{mbounce8} E. Wilson-Ewing, JCAP 1303 (2013) 026 [arXiv:1211.6269]


\bibitem{mbounce9} Yi-Fu Cai, E. Wilson-Ewing, JCAP 1403 (2014) 026 [arXiv:1402.3009 ]

\bibitem{mbounce10} J. Haro, J. Amoros, arXiv:1406.0369

\bibitem{mbounce11} J. Haro, J. Amoros, JCAP 08(2014)025 [arXiv:1403.6396 ]

\bibitem{myrza} R. Myrzakulov, L. Sebastiani, Astrophys.Space Sci. 352 (2014) 281 [arXiv:1403.0681] 

\bibitem{kaiser} D. I. Kaiser, Phys.Lett. B340 (1994) 23 [astro-ph/9405029]; D. I. Kaiser, Phys.Rev. D52 (1995) 4295 [astro-ph/9408044] 

\bibitem{sergeioikonomou} S.D. Odintsov, V.K. Oikonomou, arXiv:1410.8183

\bibitem{ref1}  G.~J.~Olmo and P.~Singh,
  JCAP 0901, 030 (2009)
  [arXiv:0806.2783 [gr-qc]].

\bibitem{ref2} S.~Baghram and S.~Rahvar,
  Phys.\ Rev.\ D  80, 124049 (2009)
  [arXiv:0912.2410 [astro-ph.CO]]

\bibitem{ref3} T.~P.~Sotiriou,
  Phys.\ Rev.\ D  79, 044035 (2009)
  [arXiv:0811.1799 [gr-qc]]
  
  \bibitem{ref4} G.~J.~Olmo and D.~Rubiera-Garcia,
  Phys.\ Lett.\ B 740 (2015) 73
  [arXiv:1405.7184 [hep-th]].
  
  \bibitem{ref5} O.~Bertolami and J.~Páramos,
  Phys.\ Rev.\ D 89, 044012 (2014)
  [arXiv:1311.5615 [gr-qc]]

\bibitem{sergeibounce} K. Bamba, A. N. Makarenko, A. N. Myagky, S. Nojiri, S. D. Odintsov, JCAP01(2014)008  [arXiv:1309.3748]


\bibitem{khoury} J. Khoury, A. Weltman, Phys. Rev. D69, 044026 (2004) [astro-ph/0309411 ]

\bibitem{sergeistarobinsky} L. Sebastiani, G. Cognola, R. Myrzakulov, S.D. Odintsov, S. Zerbini, Phys. Rev. D 89, 023518 (2014) [arXiv:1311.0744]

\bibitem{holonomyfr} X. Zhang, Y. Ma, Phys.Rev.Lett. 106, 171301 (2011) [arXiv:1101.1752]

\bibitem{mukhanov} V. Mukhanov, Physical foundations of cosmology, Cambridge, UK: Univ. Pr. (2005) 421 p; D. S. Gorbunov, V. A. Rubakov, Introduction to the theory of the early universe: Cosmological perturbations and inflationary theory, Hackensack, USA, World Scientific (2011) 489 p

\bibitem{guth} A. H. Guth, Phys.Rept. 333 (2000) 555 [astro-ph/0002156]; A. H. Guth,  J.Phys. A40 (2007) 6811 [hep-th/0702178]

\bibitem{starobinsky} A. A. Starobinsky, Phys.Lett. B91 (1980) 99

\bibitem{defnew}J. Haro and E. Elizalde, Eur. Phys. Lett. 89, 69001 (2010);  P. Dzierzak, P. Malkiewicz and W. Piechocki, Phys. Rev. D80, 104001 (2009)


\bibitem{harob} J.~Haro, A.~N.~Makarenko, A.~N.~Myagky, S.~D.~Odintsov and V.~K.~Oikonomou,
  arXiv:1506.08273 [gr-qc]

\bibitem{shad} L. Mangiarotti, G. Sardanashvily, Connections in Classical and Quantum Field Theory, (2000) World Scientific


\bibitem{kinezosvergados} Yeuk-Kwan E. Cheung, J.D. Vergados, arXiv:1410.5710

\bibitem{oikonomouvergados}V.K. Oikonomou, J.D. Vergados, Ch.C. Moustakidis, Nucl.Phys. B773 (2007) 19 [hep-ph/0612293]


\end{thebibliography}
\end{document}